\documentclass[structabstract]{aa}  
\usepackage{graphicx,xcolor,breqn}
\usepackage{txfonts}
\begin{document}
\newcommand{\pjd}[2]{\textbf{\color{green} {#1}}}
\newcommand{\pd}[2]{\textbf{\color{black} {#1}}}
\def\RL{R_{\mathcal{L}_1}}
\def\rl{$R_{\mathcal{L}_1}$}
\def\L1{\mathcal{L}_1}
\def\l1{$\mathcal{L}_1$}


\title{Mass transfer in eccentric binary systems using the \\ binary
  evolution code BINSTAR} \titlerunning{Mass transfer in eccentric
  binaries}

   \author{P.~J. Davis
          \inst{1}
          \and
          L. Siess\inst{1}
          \and
          R. Deschamps\inst{1}
          }

          \institute{Institut d'Astronomie et d'Astrophysique (IAA),
            Universit\'{e} Libre de Bruxelles, ULB CP226, Boulevard
            du Triomphe,
            B-1050 Brussels, Belgium \\
            \email{pdavis@ulb.ac.be} }


 
  \abstract
  {Studies of interacting binary systems typically assume that tidal forces
    have circularized the orbit by the time Roche lobe overflow (RLOF)
    commences. However, recent observations of ellipsoidal variables have
    challenged this assumption.}
  {We present the first calculations of mass transfer via RLOF for a binary
    system with a significant eccentricity using our new binary stellar
    evolution code. The study focuses on a 1.50+1.40 $M_{\odot}$ main
    sequence binary with an eccentricity of 0.25, and an orbital period of
    $P_{\mathrm{orb}}\approx 0.7$ d. The reaction of the stellar components
    due to mass transfer is analyzed, and the evolution of mass transfer
    during the periastron passage is compared to recent smooth particle
    hydrodynamics (SPH) simulations. The impact of asynchronism and
    non-zero eccentricity on the Roche lobe radius, and the effects of
    tidal and rotational deformation on the stars' structures, are also
    investigated.}
  {Calculations were performed using the state-of-the-art binary evolution
    code BINSTAR, which calculates simultaneously the structure of the two
    stars and the evolution of the orbital parameters.}
  {The evolution of the mass transfer rate during an orbit has a
    Gaussian-like shape, with a maximum at periastron, in qualitative
    agreement with SPH simulations. The Roche lobe radius is modified by
    the donor star's spin and the orbital eccentricity. This has a
    significant impact on both the duration and the rate of mass
    transfer. We find that below some critical rotation rate, mass transfer
    never occurs, while above some threshold, mass is transferred over the
    entire orbit. Tidal and rotational deformation of the donor star causes
    it to become over-sized, enhancing the mass transfer rate further by
    about a factor of ten, leading to non-conservative mass transfer. The
    modulation of mass transfer rate with orbital phase produces short-term
    variability in the surface luminosity and radius of each star. The
    longer-term behaviour shows, in accordance with studies of circular
    systems with radiative stars, that the donor becomes ever small and
    under-luminous, while the converse is the case for the accretor.}
  {}

  \keywords{Binaries: close -- stars: evolution -- stars: mass loss --
    methods: numerical}

   \maketitle
%

\section{Introduction}
\label{section:introduction}


Roche lobe overflow (RLOF) is of fundamental importance for a wide variety
of binary star systems (for a review see Pringle \cite{pringle}), such as
cataclysmic variables, low-mass X-ray binaries and Algols. RLOF impacts
upon both the binary orbit through the exchange (or perhaps the loss from
the system) of mass and angular momentum, and on the evolution of the
stellar components themselves, which may lead to a variety of exotic
events. Accretion onto a high-mass white dwarf, for example, may lead to
Type Ia supernovae (see Wang \& Han \cite{wang2012} for a review), or novae
events (see e.g. Warner \cite{warner}). Alternatively, a common envelope
phase is expected to result from RLOF from a red or asymptotic giant branch
star (Pazcynski \cite{paczynski}). Clearly, the outcome of binary evolution
greatly depends upon both the rate at which material is transferred, and
upon the nature of the stars.

Briefly, during RLOF, material from the donor (with mass $M_{1}$) is
channelled as a narrow stream through the inner Lagrangian point, \l1,
which falls into the gravitational potential well of the accretor (mass
$M_{2}$). Depending on the initial separation, the matter stream will
either impact directly onto the surface of the companion star, or form an
accretion disk around it. The mass transfer rate through the \l1{} point is
determined by how much the star over-fills its Roche lobe, which in turn is
dictated by the Roche lobe geometry and the donor's structure (see Ritter
\cite{ritter}; D'Antona, Mazzitelli \& Ritter \cite{dantona}).

The validity of the Roche model rests on three assumptions: that the
stellar components can be treated as point masses, that the orbit is
circular and that the donor star is rotating synchronously with the
orbit. Studies of binaries undergoing RLOF generally consider that the last
two of these assumptions hold by the time RLOF commences (e.g. Willems \&
Kolb \cite{willems2004}), the reason being that tidal forces act on very
short timescales, owing to the strong dependence of the tidal torques on
the ratio of the stellar radius with the orbital separation (Zahn
\cite{zahn}). Therefore, tides will have circularized the orbit, and
synchronized the donor star with the orbit, by the time RLOF starts.

However, the assumption of circular orbits for RLOF has been challenged by
observations. Petrova \& Orlov (\cite{petrova}) compiled a catalogue of 128
eccentric binaries which reveals that approximately 15 per cent of these
systems are semi-detached, while 5 per cent have evolved into
contact. Nicholls \& Wood (\cite{nicholls2012}) confirmed large
eccentricities, between 0.14 and 0.42, among seven ellipsoidal
variables. This is a surprising result because these systems, which are
close to filling their Roche lobes, should be nearly circular given their
short separation and the efficiency of the tidal torques.  These
observations suggest that some of these binaries will therefore fill their
Roche lobes while still possessing eccentric orbits. The idea of episodic
mass transfer at periastron corroborates the hypothesis of Jorissen et
al. (\cite{jorissen2009}) to explain the evolution of KIII-type giants on
the eccentricity-orbital period plane.

From a theoretical point of view, the assumption of a circular orbit with a
synchronously rotating donor star during RLOF has also been called into
question by Sepinsky et al. (\cite{sepinsky_07b,sepinsky_09}). They
investigated the secular orbital evolution of an eccentric binary with a
1.44 $M_{\odot}$ neutron star paired with a range of donor masses, as a
result of both conservative and non-conservative mass transfer via RLOF at
periastron. They compared the timescale for the evolution of the
eccentricity due to tides and RLOF and found that, in contrast to tides
which always act to circularize the orbit, mass transfer may either
increase or decrease the eccentricity, over timescales ranging from a few
Myr to a Hubble time. Furthermore, the timescale over which mass transfer
acts to increase the eccentricity may be shorter than the tidal timescale
which acts to decrease it. This occurs for mass ratios $q=M_{1}/M_{2} \la
0.6$, which lies in the low-mass binary regime. Similar behaviour was also
found by Sepinsky et al. (\cite{sepinsky_10}) who studied the orbital
evolution of eccentric binaries by investigating the gravitational
interaction between the matter stream and the stellar components. Hence, it
is not guaranteed that mass transfer will circularize the orbit.

The eccentric nature of the binary orbit also means that the Roche model
assumptions (i.e. where circular orbits and synchronous rotation is
assumed) no longer applies (e.g. Limber \cite{limber1963} and Savonije
\cite{savonije1978}). Sepinsky et al. (\cite{sepinsky_07a}) analysed the
effect of eccentricity and asynchronism on the Roche geometry. They found
that the Roche lobe radius for a donor star which is rotating
super-synchronously with the orbital motion at periastron will be smaller
than the radius calculated using the classical Eggleton (\cite{eggleton})
formula. The converse is true for a sub-synchronously rotating donor
star. As pointed out by Lajoie \& Sills (\cite[henceforth
LS11]{lajoie_11b}) this will impact upon both the rate and duration of mass
transfer.

The recent attempts to simulate mass transfer in eccentric binaries have
all used SPH techniques (e.g. Reg\H{o}s et al \cite{regos}, Church et al
\cite{church}, Lajoie and Sills \cite{lajoie_11a}, LS11), and did not
consider the possibility to address the problem using a binary stellar
evolution code, which can accurately model the internal structure of the
stellar components. To the best of our knowledge, only Edwards \& Pringle
(\cite{edwards}) have attempted to calculate RLOF in eccentric binaries
analytically. However, they considered a binary with a small eccentricity
of $5\times{10}^{-4}$, and only modelled the flow in the vicinity of the
\l1 point.

In this paper, we present the first simulation of mass transfer for
significantly eccentric systems using the state-of-the-art binary evolution
code BINSTAR (Siess et al. \cite{siess_13}). To facilitate comparisons with
the work by LS11, we consider their 1.50+1.40 $M_{\odot}$ binary, with an
eccentricity of $e=0.25$. In the light of the study by Sepinsky et
al. (\cite{sepinsky_07a}) we investigate the affect of asynchronous
rotation and eccentricity on the Roche lobe radius, and how this impacts on
the mass transfer rate. We also examine the response of the structure of
each star due to mass transfer, and to the deformation caused by rotation
and tides. The secular orbital evolution of the binary system is deferred
to a future study.

The paper is structured as follows. In Sect. \ref{section:method}, we
describe the BINSTAR code, how we calculate the Roche lobe radii, the mass
transfer rate, and how we account for the effects of rotation and tides on
the stellar structure. Our results are presented in Sect.
\ref{section:results}, which are discussed in Sect.
\ref{section:discussion}. We conclude with a summary of our investigation
in Sect. \ref{section:conclusions}.


\section{Computational method}
\label{section:method}


BINSTAR is an extension of the single star evolution code STAREVOL. For
further details regarding the stellar input physics, we refer the reader to
Siess, Dufour \& Forestini (\cite{siess_00}) and Siess
(\cite{siess_06,siess_07,siess_09,siess_10}), and references therein.

In the following sections, we describe the key binary input physics used in
this investigation. For a thorough description of the BINSTAR code, we
refer the reader to Siess et al. (\cite{siess_13}).

\subsection{Roche lobe radius}
\label{subsection:roche_lobe}

In an eccentric orbit, the separation between the two stellar components,
$D$, changes with time. For an orbit with a semi-major axis $a$ and an
eccentricity $e$, $D$ is found from
\begin{equation}
D=\frac{a(1-e^2)}{1+e\,\mathrm{cos}\,\nu},
\label{D}
\end{equation}
where $\nu$ is the true anomaly. Accordingly, the Roche lobe radius of the
donor star, \rl, will change along the orbit. The expression for \rl{} as
given by Eggleton (\cite{eggleton}) can be modified by replacing $a$ with
$D$, i.e.
\begin{equation}
 \RL=\frac{0.49q^{2/3}}{0.6q^{2/3}+\mathrm{ln}(1+q^{1/3})}D.
\label{R_L_egg}
\end{equation}
Henceforth, we term this the standard Roche lobe formalism.

However, this formula is only strictly valid for circular orbits and for
donors which are rotating synchronously with the orbit. We follow Sepinsky
et al. (\cite{sepinsky_07a}) and calculate \rl{} by taking into account the
eccentricity of the orbit, and any asynchronism of the donor star. The
potential in this case (normalized to the gravitational potential of the
accretor, $\frac{GM_{2}}{D}$) is given by (Sepinsky et
al. \cite{sepinsky_07a})
\begin{eqnarray}
  \Psi(x,y,z)=-\frac{q}{(x^{2}+y^{2}+z^{2})^{\frac{1}{2}}}-\frac{1}{[(x-1)^{2}+y^{2}+z^{2}]^{\frac{1}{2}}}
  \nonumber \\
  -\frac{1}{2}\mathcal{A}(1+q)(x^{2}+y^{2})+x,
\label{Psi}
\end{eqnarray}
where the $x$-axis lies along the line joining the centers of mass of the
two stars, in the direction from the donor to the accretor, the $z$-axis is
perpendicular to the plane of the orbit and is parallel to the spin angular
velocity of the donor, and the $y$-axis is perpendicular to the $x$-axis,
and completes a right-handed coordinate set. All coordinates are given in
units of $D$. In Eq. (\ref{Psi})
\begin{equation}
  \mathcal{A}=\frac{f^2(1+e)^4}{(1+e\,\mathrm{cos}\,\nu)^3}
\label{async}
\end{equation}
quantifies the degree of asynchronism and eccentricity, and $f$ is the spin
angular speed of the donor star in units of the orbital angular speed at
periastron, i.e. 
\begin{equation}
  f=\frac{\Omega_{1}}{\omega_{\mathrm{peri}}}.
  \label{f_omega}
\end{equation}
Here, the orbital angular speed, $\omega$, is given by
\begin{equation}
\omega=\frac{2\pi}{P_{\mathrm{orb}}}\frac{(1+e\,\mathrm{cos}\,\nu)^{2}}{(1-e^2)^{3/2}},
\label{Omega}
\end{equation}
where $P_{\mathrm{orb}}$ is the orbital period. Since, in general, the
potential in an eccentric system is varying with time, this will induce
oscillations of mass elements inside the star, and perturb its hydrostatic
equilibrium. However, Sepinsky et al. (\cite{sepinsky_07a}) show that such
motions are negligible if the dynamical timescale,
$\tau_{\mathrm{dyn}}=(GM_{1}/R_{1}^{3})^{-1/2}$ of the donor is much less
than the tidal timescale, $\tau_{\mathrm{tide}}=2\pi/|\omega-\Omega_{1}|$,
in which case the instantaneous shape of the star can be approximated by
the instantaneous surfaces of constant $\Psi$ (e.g. Limber
\cite{limber1963} and Savonije \cite{savonije1978}). This so-called
``first-approximation'' is valid for main sequence stars where
$P_{\mathrm{orb}}\ga 10$ hrs (Sepinsky et al. \cite{sepinsky_07a}), and
hence for our 0.7 d system.

Sepinsky et al. (\cite{sepinsky_07a}) provide fit formulae for \rl{} as a
function of $\mathcal{A}$ and $q$ for eccentric orbits and asynchronous
donors. However, because of discontinuities in some regions of the
$(\mathcal{A},q)$ parameter space (J. Sepinsky, private communication), the
volume of the Roche lobe is calculated numerically using a 32-point
Gaussian quadrature integration technique (Press et al. \cite{press1990}),
from which we derive the equivalent Roche lobe radius.

\subsection{Calculating mass transfer rates}
\label{subsection:mass_transfer_rates}

To calculate the mass transfer rates during each periastron passage, we
follow the prescription outlined in Ritter (\cite{ritter}) and Kolb \&
Ritter (\cite{kolb}).

We consider a donor star of mass $M_1$, radius $R_1$, effective
temperature $T_{\mathrm{eff},1}$, and with a mean molecular weight and
density at the photosphere, $\mu_{\mathrm{ph},1}$ and
$\rho_{\mathrm{ph},1}$ respectively. The mass transfer rate,
$\dot{M}_{1}$, in the case where material is removed from the
optically thin region of the donor's atmosphere (i.e. where the
optical depth is $\tau\la \frac{2}{3}$) is calculated using
\begin{equation}
-\dot{M}_{1}=\dot{M}_{0}\,\mathrm{exp}\left(\frac{R_{1}-R_{\L1}}{\hat{H}_{\mathrm{P}}}\right),
\label{mdot_ritter}
\end{equation}
(Ritter \cite{ritter}) where $\hat{H}_{\mathrm{P}}=H_{\mathrm{P}}/\gamma$
is the pressure scale height of the donor at the location of
\l1, which can be calculated from the pressure scale height
at the photosphere, $H_{\mathrm{P}}$, and a correction factor, $\gamma$,
which accounts for the geometry of the Roche lobe (see Appendix
\ref{section:m_transfer}). Also, $\dot{M}_{0}$ is the mass transfer rate if
the donor star exactly fills its Roche lobe, and is given by
\begin{equation}
  \dot{M}_{0} = \frac{2\pi}{\sqrt{e}}F(q)\frac{R_{\L1}^3}{GM_{1}}\left(\frac{\mathcal{R}T_{\mathrm{eff,1}}}{\mu_{\mathrm{ph,1}}}\right)^{3/2}\rho_{\mathrm{ph,1}}.
\label{mdot_0}
\end{equation}
Here, $\mathcal R$ is the ideal gas constant, $G$ is the gravitational
constant and $F(q)$ is a function of $q$, and is determined from the area
of the equipotential surface which intersects with the \l1{}
point:
\begin{equation}
F(q)=q\,\beta^{-3}\left\{ g(q)\left[g(q)-1-q\right]\right\} ^{-1/2}.
\label{Fq}
\end{equation}
In turn, $\beta=\frac{\RL}{D}$, and $g(q)$ is given by
\begin{equation}
g(q)=\frac{q}{x_{\L1}^3}+\frac{1}{(1-x_{\L1})^3},
\label{gq}
\end{equation}
where $x_{\L1}$ is the distance from the centre of mass of the donor star
to the \l1 point, in units of $D$. For the Roche model, $x_{\L1}$ can be
determined by numerically solving
\begin{equation}
  \frac{q}{x_{\L1}^{2}}-\frac{1}{(x_{\L1}-1)^2}-\mathcal{A}x_{\L1}(1+q)+1=0
\label{x_L}
\end{equation}
(Sepinsky et al. \cite{sepinsky_07a}), with $\mathcal{A}=1$ for
  synchronously rotating stars in circular orbits.

  If the donor is significantly overflowing its Roche lobe, or if the
  donor's radius and Roche lobe radius are not evolving in step, then
  Eq. (\ref{mdot_ritter}) is no longer valid. Instead, mass is also lost
  from the optically thick layers of the star (where $\tau \ga 2/3$). The
  mass transfer rate in this case becomes (see also Deloye et
  al. \cite{deloye2007})
\begin{eqnarray}
  -\dot{M}_{1}& = &\dot{M}_{0}+2\pi
  F(q)\frac{R_{\L1}^3}{GM_{1}}\times \nonumber \\ & & \int^{R_{\mathrm{ph}}}_{R_{\L1}}\Gamma_{1}^{1/2}\left(\frac{2}{\Gamma_{1}+1}\right)^{\frac{\Gamma_{1}+1}{2(\Gamma_{1}-1)}}\frac{GM_{r}(P\rho)^{\frac{1}{2}}}{r^2}\,\mathrm{d}r,
\label{mdot_kolb}
\end{eqnarray}
where $P$, $T$, $\mu$ and $M_{r}$ is the pressure, temperature, mean
molecular weight and the mass of the donor star respectively at the radial
coordinate $r$, and
$\Gamma_{1}=(\mathrm{d}\,\mathrm{ln}P/\mathrm{d}\,\mathrm{ln}\rho)_{\mathrm{ad}}$
is the adiabatic exponent. The integral in Eq. (\ref{mdot_kolb}) is
evaluated numerically from the \l1 point to the photosphere (subscript
`ph').

If the donor star is rotating asynchronously with the orbit, and the
eccentricity is non-zero, Eq. (\ref{Fq}) is modified according to (see
Appendix \ref{section:m_transfer})
\begin{equation}
F^{*}(q,\mathcal{A})=q\beta^{-3}\left\{g(q)\left[g(q)-\mathcal{A}-q\mathcal{A}\right]\right\}^{-\frac{1}{2}},
\label{F_star}
\end{equation}
where the asterisk makes the distinction from Eq. (\ref{Fq}).

\subsection{Initial model}
\label{sub:initial_model}

\subsubsection{The binary model}

In order to compare our results with the SPH simulations of LS11, we
consider their 1.50$+$1.40 $M_{\odot}$ main sequence binary system
configuration. They construct their SPH models from the theoretical density
profiles of a 1.50 $M_{\odot}$ donor star with a 1.40 $M_{\odot}$ accreting
companion, calculated from their stellar evolution code, YREC (Guenther et
al. \cite{guenther_1992}; see LS11 for further details).

Their stars have a metallicity of $Z=0.001$, and an age of approximately
1.3 Gyr, representative of binary systems populating old open
clusters. They use a mixing length parameter $\alpha_{\mathrm{MLT}}=1.71$,
with no convective overshooting or rotation (A. Sills, private
communication). With their input physics, we find that the donor and
accretor have a radius of approximately 1.4 and 1.2 $R_{\odot}$
respectively, in agreement with LS11. We also follow LS11 and synchronize
the spin angular velocity of the stars to the orbital angular speed at
apastron, yielding $f\approx 0.36$. The surface angular speed of the
stellar components is then $\Omega_{1}=\Omega_{2}\approx 6\times{10}^{-5}$
s$^{-1}$. For simplicity, we assume that the stars rotate as solid bodies,
since the treatment of differential rotation is beyond the scope of the
paper.

Pertinent to this study, we find that each star possesses a thin surface
convective envelope, due to the first ionization of hydrogen. Their radial
extents are relatively small (approximately 0.0014 R$_\odot$ for the
1.50M$_\odot$ donor, and $0.002$ $R_{\odot}$ for the 1.40M$_\odot$
accretor), and their masses negligible, but they are crucial for
understanding the response of the stellar structure to mass changes (see
Section \ref{sec:reaction}). Each star also has two additional convection
zones\footnote{For the accretor, the convection zones due to ionized
  hydrogen and the first ionization of helium are merged into a single
  zone.} associated with the first and second ionization of helium, but
they play a secondary role. Henceforth, we denote the zone associated with
the ionization of hydrogen as the surface convection zone. The upper layers
of this convective region are super-adiabatic and energy transport via
convection becomes increasingly inefficient, until all the flux is
transported purely by radiation in the atmosphere. To aid numerical
stability, the number of shells in the donor and accretor are kept constant
and are equal to 818 and 847 respectively.

\subsubsection{Integrating the orbit}

With an eccentricity, $e=0.25$ and a semi-major axis, $a=4.80$ $R_{\odot}$,
the orbital separation at periastron is $r_{\mathrm{peri}}=3.60$
$R_{\odot}$, and the orbital period $P_{\mathrm{orb}}\approx 0.7$ days.

To resolve numerically the orbital motion, we impose the evolutionary
timestep $\Delta t$ not to exceed some fraction,
$\mathcal{F}_{\mathrm{orb}}$, of the time it takes for the stars to travel
the circumference $2\pi a$, given a value $v_{\mathrm{orb}}$ of the orbital
speed at that orbital phase i.e.
\begin{equation}
  \Delta t=\mathcal{F}_{\mathrm{orb}}\left(\frac{2\pi a}{v_{\mathrm{orb}}}\right),
  \label{delta_t}
\end{equation}
with
\begin{equation}
  v_{\mathrm{orb}}=\sqrt{G(M_{1}+M_{2})\left(\frac{2}{D}-\frac{1}{a}\right)}\,\,.
  \label{vorb}
\end{equation}
For our simulations, we use $\mathcal{F}_{\mathrm{orb}}=5\times{10}^{-3}$,
which give us timesteps of approximately $10^{-6}$ yr.

\subsection{Treatment of mass loss and mass gain}

In the surface layers of the star, variations in the luminosity result from
the release of gravo-thermal energy per unit time and per unit mass
\begin{equation}
  \varepsilon_{\mathrm{grav}}=-T\left(\frac{\partial s}{\partial t}\right)_{m},
  \label{egrav}
\end{equation}
where $s$ is the specific entropy, and the subscript $m$ denotes that the
derivative is evaluated at a fixed mass coordinate. In the situation where
mass is lost or gained by the star, mass and time can no longer be
considered as independent variables. Following Neo, Miyaji \& Nomoto
(\cite{neo1977}) we use a pseudo-Lagrangian variable, $\tilde{q}_{i}$, for
each star $i=1,\,2$, which is defined as
\begin{equation}
  \tilde{q}_{i}=\frac{m_{i}-M_{i}^{\prime}}{M_{i}(t)-M_{i}^{\prime}},
\end{equation}
where $M_{i}^{\prime}$ is the mass coordinate above which mass is lost or
gained, $m_{i}$ is a mass coordinate located at $m_{i} \ge M_{i}^{\prime}$
and $M_{i}(t)$ is the stellar mass at time $t$. In those layers affected by
a change in mass, $\tilde{q}_{i}$ varies between 0 and 1. For the donor, we
take the value of $M_{1}^{\prime}$ to coincide with the location of the
Roche lobe radius inside the star. We ensure that the time-step, $\Delta
t$, is small enough such that $|\dot{M}_{1}|\Delta
t<M_{1}-M_{1}^{\prime}$. For the accretor, mass is deposited uniformly
above the mass coordinate corresponding to a fraction $f_{\mathrm{accr}}$
of the accretor's mass. Hence, $M_{2}^{\prime}=f_{\mathrm{accr}}M_{2}$,
where we arbitrarily set $f_{\mathrm{accr}}=0.9$. We find that the response
of the accretor to mass gain is independent of the value of
$f_{\mathrm{accr}}$.

In this scheme, Eq. (\ref{egrav}) can be re-cast as (see Neo et al.
\cite{neo1977}, Fujimoto \& Iben \cite{fujimoto1989})
\begin{eqnarray}
  \varepsilon_{\mathrm{grav}}& = &-T\left(\frac{\partial s}{\partial
      t}\right)_{\tilde{q}_{i}}+T\left(\frac{\partial s}{\partial
      \mathrm{ln}\tilde{q}_{i}}\right)_{t}\frac{\partial \mathrm{ln}M_{i}}{\partial t}, \nonumber
  \\
                        & = & \varepsilon_{\mathrm{grav}}^{(\mathrm{nh})}+\varepsilon_{\mathrm{grav}}^{(\mathrm{h})}, 
    \label{grav_2}
\end{eqnarray}
where the subscripts $\tilde{q}_{i}$ and $t$ indicate that the derivatives
are to be evaluated at constant $\tilde{q}_{i}$ and $t$ respectively. The
first and second terms on the right hand side of Eq. (\ref{grav_2})
correspond to the non-homologous,
$\varepsilon_{\mathrm{grav}}^{(\mathrm{nh})}$, and homologous,
$\varepsilon_{\mathrm{grav}}^{(\mathrm{h})}$, gravo-thermal energy
generation rates, respectively.

The gravo-thermal luminosity, $L_{\mathrm{grav}}$, is then the sum of the
non-homologous, $L_{\mathrm{grav}}^{(\mathrm{nh})}$, and homologous,
$L_{\mathrm{grav}}^{(\mathrm{h})}$, contributions given by
\begin{eqnarray}
  L_{\mathrm{grav}} & = &
  L_{\mathrm{grav}}^{(\mathrm{nh})}+L_{\mathrm{grav}}^{(\mathrm{h})}, \nonumber \\
                   & = &
                    \int^{M_{i}}_{M_{i}^{\prime}}\varepsilon_{\mathrm{grav}}^{(\mathrm{nh})}\,\mathrm{d}m
                    +
                    \int^{M_{i}}_{M_{i}^{\prime}}\varepsilon_{\mathrm{grav}}^{(\mathrm{h})}\,\mathrm{d}m.
 \label{Lgrav}
\end{eqnarray}

It is unclear what fraction, $0\leq \alpha_{\mathrm{acc}}\leq 1$, of the
accretion luminosity, $L_{\mathrm{acc}}$, is imparted to the stellar
layers, and what fraction is radiated away. For the case of direct-impact
accretion Ulrich \& Burger (\cite{ulrich}) argue that, due to the small
fraction of the accretor's surface that is covered by the hot-spot, the
energy dissipated from the shock region will have a negligible effect on
the internal structure of the star. For disc accretion, it is uncertain
what fraction of the luminosity emitted from the star-disc boundary layer
is absorbed by the accretor (e.g. Siess, Forestini \& Bertout
\cite{siess_97}). As shown in Sect. \ref{sec:accretion}, however, accretion
occurs via direct impact. Hence, following e.g.  Kippenhahn \&
Meyer-Hofmeister (\cite{kippenhahn_76}), Tout et al. (\cite{tout}), Braun
\& Langer (\cite{braun}), we assume that the accreted mass has the same
specific entropy as the shell in which it is deposited.

For the models presented in Sections \ref{sub:mass_transfer} and
\ref{sec:reaction}, we also assume that mass transfer is fully
conservative, i.e. $\beta=|\dot{M}_{2}/\dot{M}_{1}|=1$. We investigate the
possibility of relaxing this assumption further in Section
\ref{sec:accretion}.

\subsection{Effects of tides and rotation on the stellar structure}
\label{sec:tides}

In order to account for the deformation of the stellar structure caused by
tidal and rotational forces, we implemented the prescription described by
Landin, Mendez \& Vaz (\cite{landin}), and Song, Zhong \& Lu
(\cite{song}). Their method employs the technique developed by Kippenhahn
\& Thomas (\cite[henceforth KT70]{kippenhahn}), and improved by Endal \&
Sofia (\cite{endal}), to quantify the distortion of the star in a
1-dimensional stellar evolution code.

Consider an equipotential, $\Psi$, of surface $S_{\Psi}$ and volume
$V_{\Psi}$, which encloses a mass $M_{\Psi}$. Following KT70, the stellar
structure equations are modified by applying the correction factors
$f_{\mathrm{P}}$ and $f_{\mathrm{T}}$, which are respectively given by

\begin{equation}
f_{\mathrm{P}}=\frac{4\pi
  r_{\Psi}^{4}}{GM_{\Psi}}\frac{1}{S_{\Psi}\langle{g^{-1}}\rangle},
\label{f_p}
\end{equation}
and
\begin{equation}
  f_{\mathrm{T}}=\left(\frac{4\pi
      r_{\Psi}^{2}}{S_{\Psi}}\right)^{2}\frac{1}{\langle{g}\rangle \langle{g^{-1}}\rangle},
\label{f_t}
\end{equation}
where $r_{\Psi}$ is the radius of a sphere with volume $V_{\Psi}$. The
effective gravities, averaged over $S_{\Psi}$, are given by
\begin{equation}
  \langle g \rangle=\frac{1}{S_{\Psi}}\int_{0}^{\pi}\int_{0}^{2\pi}g(r,\theta,\phi)\,
  r^{2}\sin\theta\,\mathrm{d}\theta\,\mathrm{d}\phi,
\label{g_eff_mean}
\end{equation}
where the local gravity, $g(r,\theta,\phi)$, is obtained by differentiating
the potential at the considered location, $P(r,\theta,\phi)$, i.e.
\begin{equation}
g(r,\theta,\phi)=\left[\left(\frac{\partial\Psi}{\partial
      r}\right)^{2}+\left(\frac{1}{r}\frac{\partial\Psi}{\partial\theta}\right)^{2}+\left(\frac{1}{r\sin\theta}\frac{\partial\Psi}{\partial\phi}\right)^{2}\right]^{1/2}.
\label{g_eff}
\end{equation}

\begin{figure}
  \begin{center}
    \includegraphics[trim = 0mm 0mm 0mm
    10mm,clip,scale=0.45]{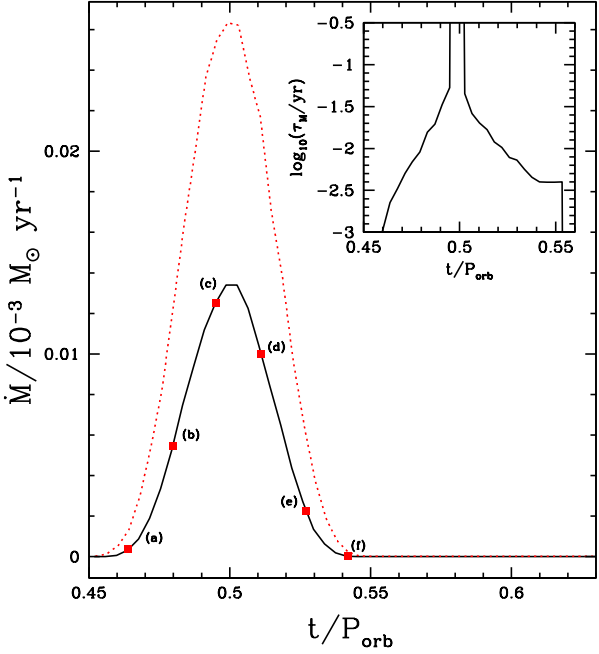}
    \caption{Mass loss rate, $\dot{M}_{1}$ during the first periastron
      passage as a function of time since apastron, in units of
      $P_{\mathrm{orb}}$ when the effects of tides and rotation on the
      stellar structure are ignored (black, solid curve) or included (red,
      dotted curve). The red squares labelled (a) to (f) on the black curve
      indicate different moments in time, and correspond to the panels
      labelled (a) to (f) in Figs. \ref{Legravs} and \ref{Legravs2}. The
      inset shows the evolution of the timescale over which the mass
      transfer rate changes, $\tau_{\dot{M}}=|\dot{M}_{1}/\ddot{M}_{1}|$
      for the non-distorted model.}
  \label{mtransfer_01} 
  \end{center}
\end{figure}

In this expression, the potential consists of four contributions; the
spherically symmetric part of the gravitational potential,
$\Psi_{\mathrm{grav}}$, the cylindrically symmetric potential due to
rotation, $\Psi_{\mathrm{rot}}$, the potential due to tides
$\Psi_{\mathrm{tide}}$ and the (non-symmetric) gravitational potential due
to the distortion of the star resulting from rotation,
$\Psi_{\mathrm{dist,tide}}$ and tides $\Psi_{\mathrm{dist,rot}}$, and can
be written
\begin{eqnarray}
  \Psi(r,\theta,\phi)& =& \Psi_{\mathrm{grav}}+\Psi_{\mathrm{rot}}+\Psi_{\mathrm{tide}}+\Psi_{\mathrm{dist,rot}}+\Psi_{\mathrm{dist,tide}},\nonumber
  \\
  & = &
  \frac{GM_{\Psi}}{r}+\frac{1}{2}\Omega_{1}^{2}r^{2}\sin^{2}\theta+\frac{GM_{\mathrm{2}}}{D}\left[1+\sum_{j=2}^{4}\left(\frac{r_{0}}{D}\right)^{j}\mathcal{P}_{j}(\lambda)\right]\nonumber
  \\ & & -\frac{4\pi}{3r^{3}}\mathcal{P}_{2}(\cos\theta)\,\mathcal{J}+4\pi GM_{\mathrm{2}}\sum_{j=2}^{4}\frac{\mathcal{P}_{j}(\lambda)}{(r\,
    D)^{j+1}}\,\mathcal{I}_{j},
     \label{psi_rot_tide}
\end{eqnarray}
where $\lambda\equiv\sin\theta\,\cos\phi$, $r$ is the distance from the
centre of the star to $P(r,\theta,\phi)$ and $\mathcal{P}_j$ is the $j$th
Legendre polynomial. The integrals $\mathcal{J}$ and $\mathcal{I}_{j}$ are
respectively given by
\begin{equation}
\mathcal{J}=\int_{0}^{r_{0}}\rho\frac{r_{0}^{\prime7}}{M_{\Psi}}\frac{5+\eta_{2}}{2+\eta_{2}}\,\Omega_{1}^{2}(r_{0}^{\prime})\,\mathrm{d}r_{0}^{\prime}
\label{int_J}
\end{equation}
and
\begin{equation}
\mathcal{I}_{j}=\int_{0}^{r_{0}}\rho\frac{r_{0}^{\prime2j+3}}{M_{\Psi}}\frac{j+3+\eta_{j}}{j+\eta_{j}}\,\mathrm{d}r_{0}^{\prime}
\label{int_I}
\end{equation}
for $j=2,3,4$ (Kopal \cite{kopal}). In Eqs. (\ref{psi_rot_tide}) to
(\ref{int_I}), $r_0$ is the mean radius of the considered equipotential
surface, and $\eta_{j}$ is determined numerically by solving the Radau
equation
\begin{equation}
r_{0}\frac{d\eta_{j}}{dr_{0}}+6\frac{\rho(r_{0})}{\langle\rho(r_{0})\rangle}(\eta_{j}+1)+\eta_{j}(\eta_{j}-1)=j(j+1),
\label{radau}
\end{equation}
using the boundary condition $\eta_{j}(0)=j-2$ (see Landin et
al. \cite{landin} and Song et al. \cite{song} for details).

Note that in their study, Song et al. (\cite{song}) only use the above
prescription to the interior of the star, while an analytical approximation
is applied at the surface (H.~F. Song: private communication). Furthermore,
they neglect terms higher than second order (i.e. $j>2$). In doing so, they
predict equal values for the effective gravity at the location facing the
companion and on the opposite side (see their Fig. 2). On the contrary, we
find that tidal deformation at the surface closer to the companion is
larger, giving a lower effective gravity here, because in our approach we
take into account the higher-order terms.
 


\section{Results}
\label{section:results}

In the next Section, we present the mass loss rate from the donor
calculated using the standard Roche lobe formalism (Eq.~\ref{R_L_egg}), and
in Sect. \ref{sec:reaction} we analyze the structural response of the
stellar components due to the mass exchange. The impact of tides and
rotation on the structure of the so-called `distorted models' is also
discussed. In Section \ref{sec:accretion}, we examine to what extent this
evolution remains conservative, while in Section \ref{sub:asynch} the
effects of asynchronism and eccentricity on the Roche lobe radius and mass
loss rate are investigated.


\subsection{Mass transfer rates}
\label{sub:mass_transfer}

\begin{figure}
  \begin{center}
    \includegraphics[trim = 0mm 230mm 500mm
    0mm,clip,scale=0.45]{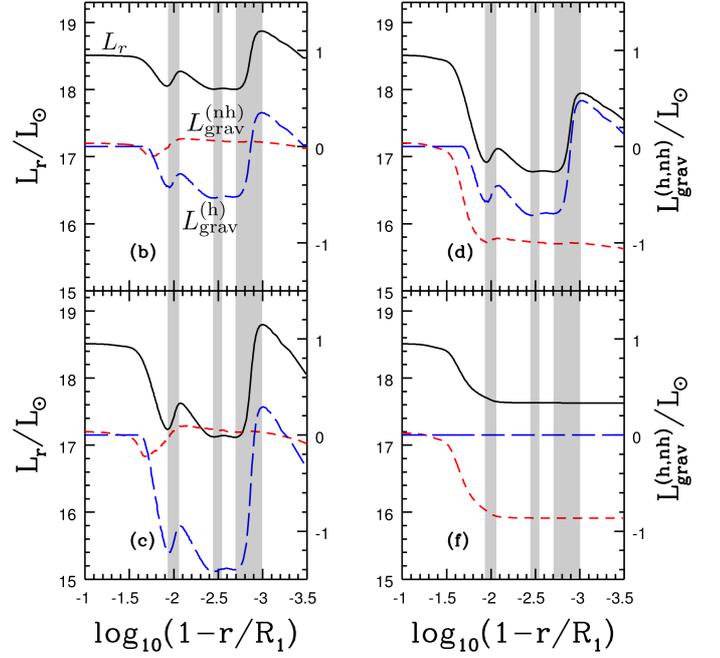}
    \caption{Stellar luminosity profile, $L_{r}$ (black solid curve, left
      axis), and contributions from the homologous,
      $L_{\mathrm{grav}}^{(\mathrm{h})}$ (blue, short-dashed curve, right
      axis) and the non-homologous, $L_{\mathrm{grav}}^{(\mathrm{nh})}$
      (red, long-dashed curve, right axis) terms for the non-distorted
      donor star at different moments in time, specified in the bottom
      left-hand corner of each panel (see Fig. \ref{mtransfer_01}). The
      shaded regions indicate convection zones.}
    \label{Legravs}
  \end{center}
\end{figure}

As we can see from Fig.~\ref{mtransfer_01}, RLOF commences just before
periastron, and ceases just after, with a maximum mass transfer rate at
closest approach of $\dot{M}_{1}\approx 1.3\times 10^{-5}$ $M_{\odot}$
yr$^{-1}$ for the non-distorted model. Moving towards periastron, the
donor's Roche lobe radius shrinks as the separation between the two stars
decreases (Eq. \ref{R_L_egg}). The amount that the donor star overfills its
critical Roche surface (hereafter termed the overfilling factor)
consequently rises, and $\dot{M}_{1}$ increases (Eqs. \ref{mdot_ritter} and
\ref{mdot_kolb}). After reaching a maximum at periastron, both the degree
of RLOF and $\dot{M}_{1}$ decline. This produces a Gaussian-like shape in
the time evolution of $\dot{M}_{1}$, where the duration of RLOF is
approximately 11 per cent of the orbital period.

\begin{figure*}
  \begin{minipage}{185mm}
    \begin{center}
    \includegraphics[trim = 0mm 0mm 0mm
    5mm,clip,scale=0.45]{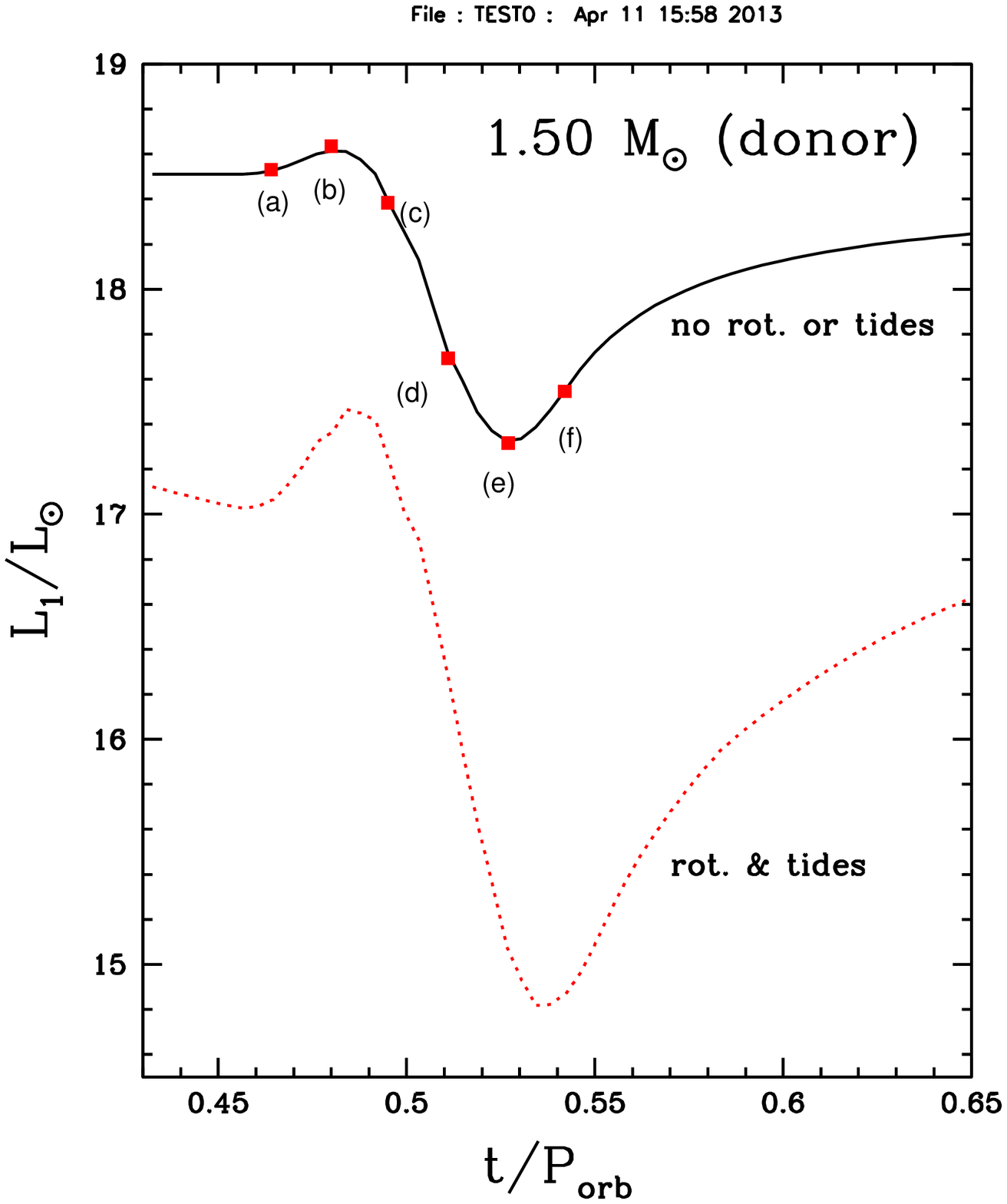}
    \includegraphics[trim = 0mm 0mm 0mm
    5mm,clip,scale=0.45]{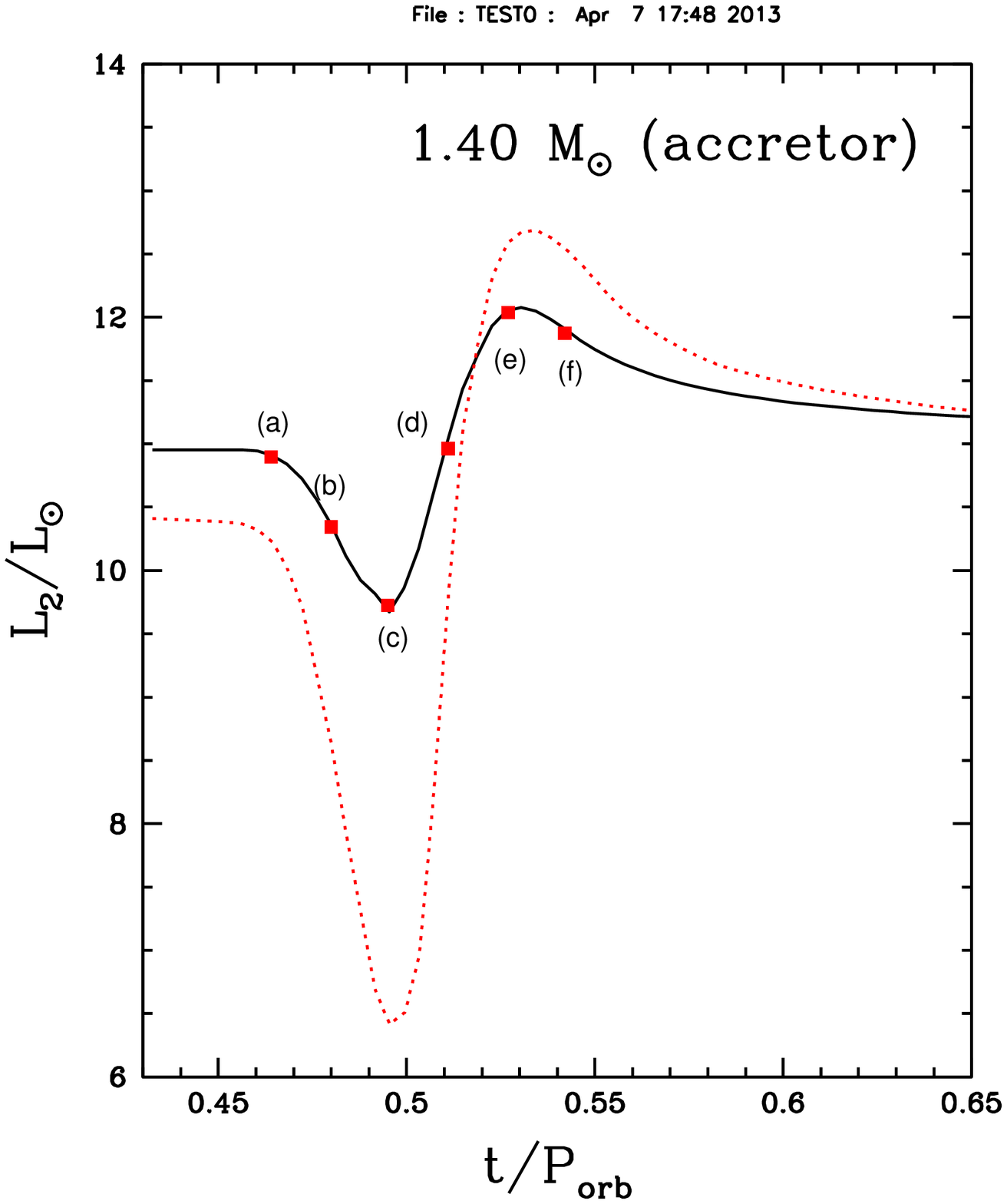}
    \caption{Evolution of the surface luminosity for the donor, $L_{1}$
      (left panel), and of the accretor, $L_{2}$ (right panel) during mass
      transfer at the first periastron passage. The meaning of the lines
      are the same as for Fig.~\ref{mtransfer_01}. Points labelled (a) to
      (f) correspond to the points labelled (a) to (f) in
      Fig.~\ref{mtransfer_01}.}
    \label{rad_lum_01} 
   \end{center}
  \end{minipage}
\end{figure*}

For the distorted model, $\dot{M}_{1}$ (at a given phase) is higher than
for the non-distorted case (dashed, red curve in Fig.~\ref{mtransfer_01}),
and at periastron peaks at approximately $2.6\times{10}^{-5}$ $M_{\odot}$
yr$^{-1}$. The reason stems from the fact that the combined effects of
tides and rotation reduce the effective surface gravity $\langle g
\rangle$, increasing the donor's radius (a relative increase of about 0.4
per cent compared to the non-distorted model), and thus the overfilling
factor at any orbital phase. The effect of increased stellar radius has
also been reported within the SPH simulations of LS11 and Renvoiz\'{e} et
al. (\cite{renvoize}) (see also Uryu \& Eriguchi \cite{uryu}, who use a
different computational technique). The small increase in the donor's
radius has a significant impact on $\dot{M}_{1}$ due to its sensitivity on
the overfilling factor, but marginally affects the duration of mass
transfer, increasing it to about 12 per cent.

Such a modulation of the mass transfer rate with orbital phase was
suggested to account for a change in the speed of the bipolar outflows
emanating from the accretor of HD 44179 within the Red Rectangle. In this
paradigm, the velocity of the ejected material is maximum at periastron,
and minimum at apastron (Witt et al. \cite{witt}). Furthermore, variations
of the X-ray luminosity of the intermediate-mass black hole HLX-1 (Lasota
et al. \cite{lasota}) and of the optical and ultra-violet luminosity of the
symbiotic system BX Mon (Leibowitz et al. \cite{leibowitz}) have also been
attributed to such a modulation of the mass transfer rate.

\begin{figure*}
  \begin{minipage}{185mm}
    \begin{center}
    \includegraphics[trim = 0mm 0mm 0mm
    10mm,clip,scale=0.45]{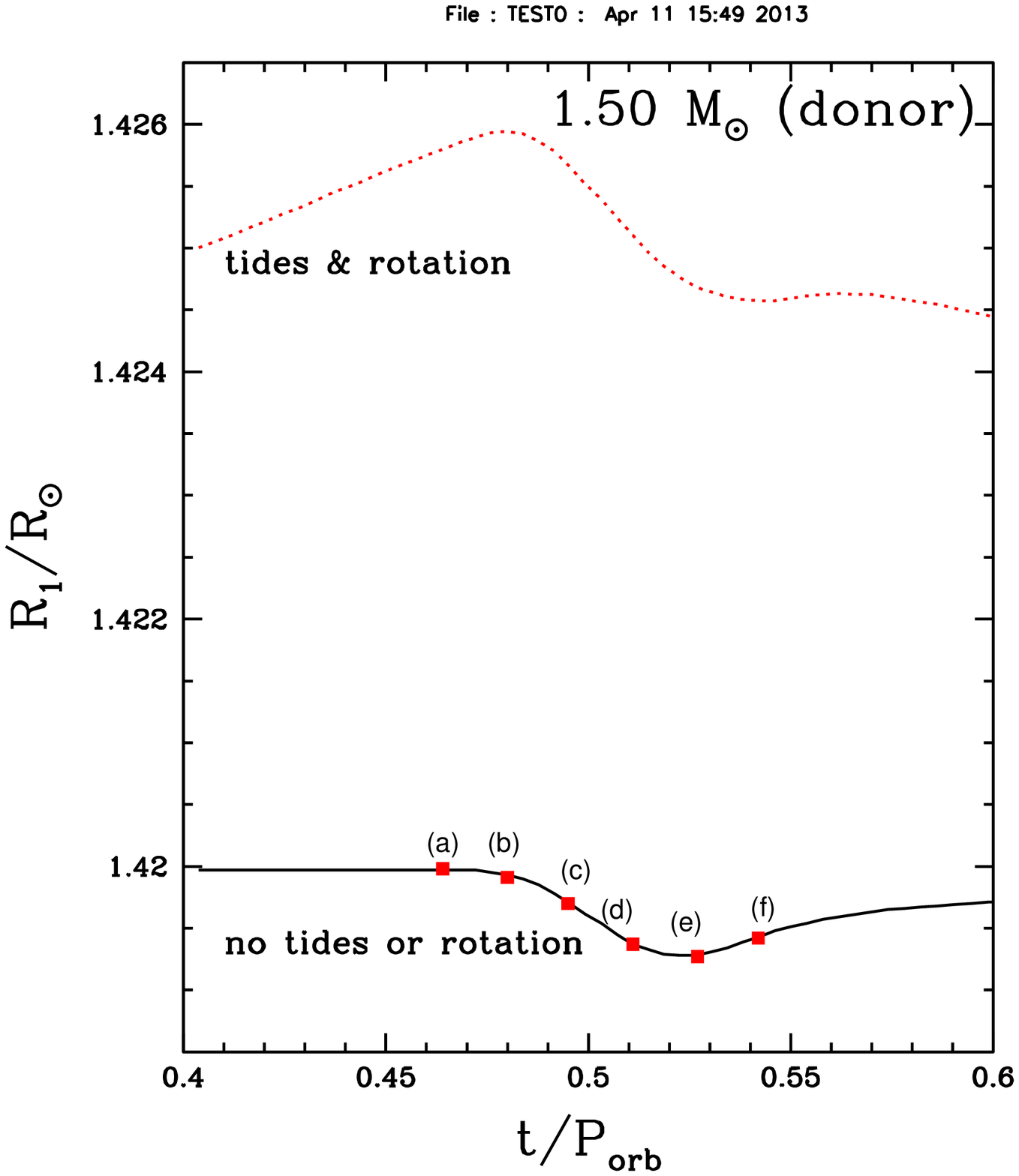}
    \includegraphics[trim = 0mm 0mm 0mm
    10mm,clip,scale=0.45]{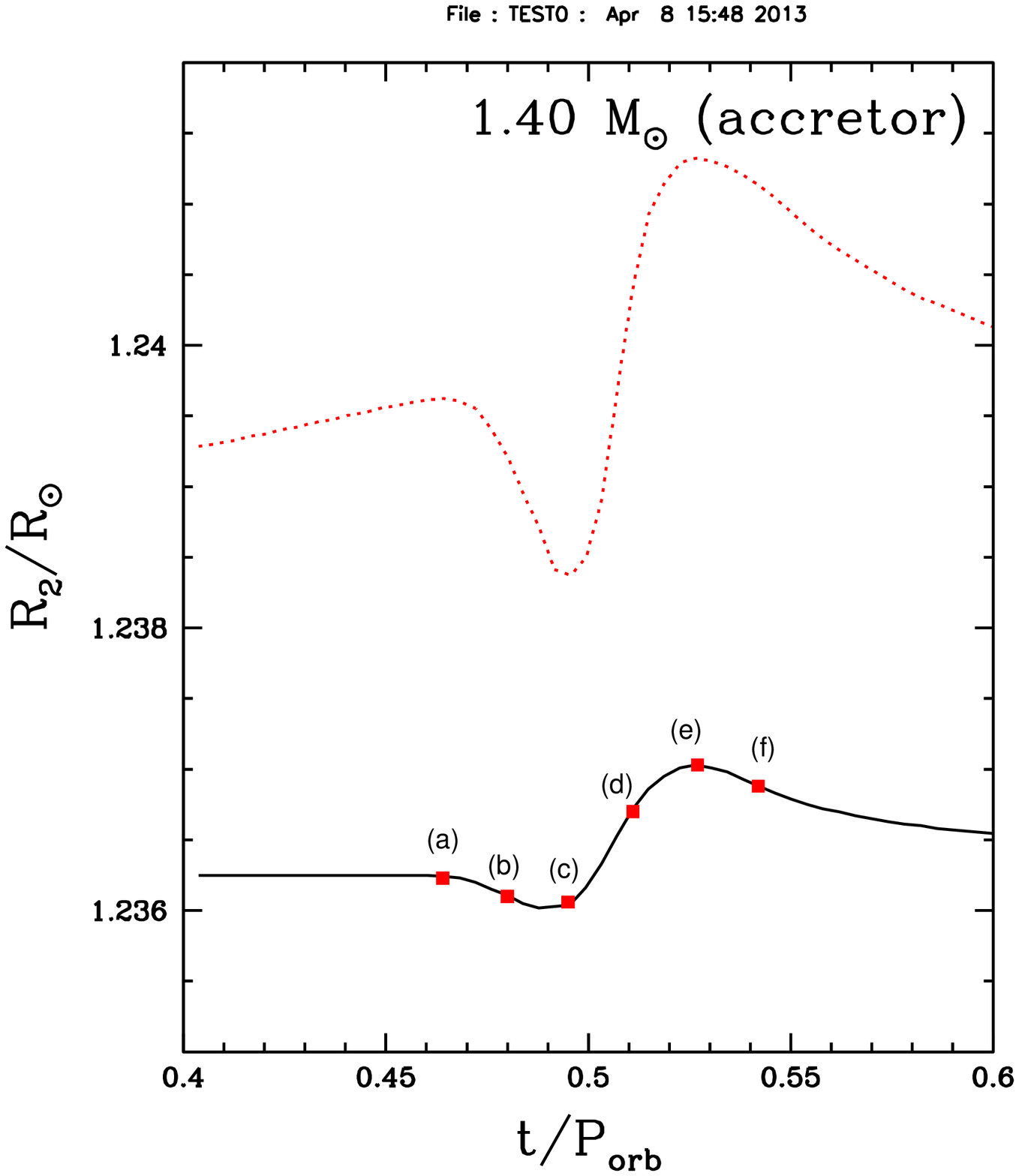}
    \caption{Similar to Fig.~\ref{rad_lum_01}, but now depicting the
      evolution of the radius for the donor $R_{1}$ (left panel) and the
      accretor $R_{2}$ (right panel) during the first periastron passage.}
    \label{rad1} 
  \end{center}
\end{minipage}
\end{figure*}

\subsection{The reaction of the stars}
\label{sec:reaction}

In contrast to circular orbits, mass transfer occurs periodically during
each periastron passage, and changes with the orbital phase
(Fig.~\ref{mtransfer_01}), which gives rise to a corresponding short-term
variability of the radius and luminosity of each star. The response of the
donor and accretor during mass transfer is discussed in
Sects. \ref{sub:donor_reaction} and \ref{sub:reaction_accretor},
respectively.

\subsubsection{The donor star}
\label{sub:donor_reaction}

In accordance with e.g. Webbink (\cite{webbink77a}) and Neo et
al. (\cite{neo1977}), the action of removing mass from radiative layers
(where the entropy gradient is positive) absorbs energy ($\partial
L_{\mathrm{grav}}^{(\mathrm{h})}/\partial{r}<0$; see blue, short-dashed
curves in Fig.~\ref{Legravs}). Conversely, in the super-adiabatic region
(right-most shaded area in Fig.~\ref{Legravs}) mass loss causes a slight
increase in $L_{r}$ because in those layers $\partial
L_{\mathrm{grav}}^{(\mathrm{h})}/\partial{r}>0$.

As discussed in Webbink (\cite{webbink76}), a changing mass transfer rate
will perturb the thermal structure of the star over a timescale
$\tau_{\dot{M}}=|\dot{M}_{1}/\ddot{M}_{1}|$. Initially, $\tau_{\dot{M}}$ is
so short (inset of Fig.~\ref{mtransfer_01}) that only the outermost surface
layers, which consist of the surface convection zone, can restore thermal
equilibrium on a timescale $\tau_{\mathrm{th}}\ll\tau_{\dot{M}}$. In these
superadiabatic layers [which encompass about 3 per cent of the donor radius
at point (b)], $\epsilon_{\mathrm{grav}}^{(\mathrm{nh})}\approx 0$ and the
luminosity increases due to the dominant contribution from
$\epsilon_{\mathrm{grav}}^{(\mathrm{h})}$ [panel (b), Fig.~\ref{Legravs}].

Subsequently, as $\dot{M}_{1}$ increases towards periastron, an
ever-growing deficit in $L_{\mathrm{grav}}^{(\mathrm{h})}(\propto
\dot{M}_{1})$ within the radiative layers develops. In parallel, as
$\tau_{\dot{M}}$ is also increasing, the condition $\tau_{\mathrm{th}}<
\tau_{\dot{M}}$ moves deeper [encompassing about 10 per cent of the donor's
radius at point (c)]. Because $L_{\mathrm{grav}}^{(\mathrm{h})}$ dominates
over a larger fraction of the radiative envelope, the luminosity deficit
grows in these layers (black curves in Fig.~\ref{Legravs}) and entails an
over-all reduction of the star's surface luminosity $L_{1}$ [e.g. point
(c), left panel of Fig.~\ref{rad_lum_01}] and radius $R_{1}$ [left panel,
point (c) on black curve of Fig.~\ref{rad1}], as attested by
$L_{\mathrm{grav}}^{(\mathrm{nh})}>0$ (red long-dashed curves, panels (b)
and (c) in Fig.~\ref{Legravs}).

Away from periastron, both $|L_{\mathrm{grav}}^{(\mathrm{h})}|$ and
$\tau_{\dot{M}}$ decline. The former will still cause $R_{1}$ and $L_{1}$
to shrink [Figs.  \ref{rad_lum_01} and \ref{rad1}, e.g. point (d)] but at a
slower rate. The latter will give rise to a decrease in
$L_{\mathrm{grav}}^{(\mathrm{nh})}$ due to decompression of the deep
radiative layers where $\tau_{\dot{M}}<\tau_{\mathrm{th}}$, and eventually
$L_{\mathrm{grav}}^{(\mathrm{nh})}$ becomes negative [panel (d) of
Fig.~\ref{Legravs}]. Once mass loss shuts off, the radiative layers expand
as energy flows from the interior to fill the luminosity deficit and
restore thermal equilibrium [Fig.~\ref{Legravs}, panel (f)], causing a
corresponding rise in $L_{1}$ and $R_{1}$ [e.g. point (f) in
Figs. \ref{rad_lum_01} and \ref{rad1}].

By the time the donor starts overfilling its Roche lobe at the second
periastron passage, the star has not yet fully recovered thermal
equilibrium and $R_{1}$ and $L_{1}$ are slightly smaller than at the
beginning of the simulation. Subsequent mass transfer episodes perturb the
structure further, and $R_{1}$ and $L_{1}$ become ever smaller at each new
mass transfer episode (black curve, Fig. \ref{L1_L2}). Despite the
short-term reaction of the donor during periastron, the aforementioned
longer-term behaviour is in accordance with studies of predominantly
radiative stars in circular orbits (e.g. Webbink
\cite{webbink77a,webbink77b}, Siess et al. \cite{siess_13}).


For the distorted model (red, dotted curves in Figs. \ref{rad_lum_01} and
\ref{rad1}), the donor is over-sized and under-luminous compared to the
non-distorted model for all orbital phases. The evolution of $R_{1}$ and
$L_{1}$ during the periastron passage are qualitatively the same as for the
non-distorted model, but for two main differences. Firstly, before mass
transfer starts, $R_{1}$ rises while $L_{1}$ declines. This is because, as
the stellar separation shrinks, the surface gravity decreases due to the
strengthening of the tidal interaction terms on the right hand side of
Eq. (\ref{psi_rot_tide}). In turn, the stars' structure will be less
compact, and the reverse process occurs once the stellar separation
increases. Secondly, the evolution of $R_{1}$ and $L_{1}$ during mass
transfer is perturbed more significantly, due to the larger mass transfer
rates obtained in the tidally distorted case, as discussed in Section
\ref{sub:mass_transfer}.

\subsubsection{The accreting star}
\label{sub:reaction_accretor}

As with the donor, the reaction of the accretor is initially dictated by
the surface convection zone, but because the extent of that region is about
twice that of the donor's, it governs the reaction over a longer duration
[almost until periastron, see panels (b) and (c) in
Fig.~\ref{Legravs2}\footnote{The values of
  $L_{\mathrm{grav}}^{(\mathrm{nh})}$ and
  $L_{\mathrm{grav}}^{(\mathrm{h})}$ may seem large compared to the surface
  luminosity but what is physically relevant is the sum of these two
  contributions which, as previously stated, is not affected by the choice
  of $f_{\mathrm{accr}}$.}]. The response of the accretor is opposite to
that of the donor star; mass addition to the convective layers causes the
surface luminosity, $L_{2}$, and radius, $R_{2}$, to shrink [points (b) to
(c) in right panels of Fig.~\ref{rad_lum_01} and Fig.~\ref{rad1}].

As the stars move away form periastron, the compression decelerates as
$\tau_{\dot{M}}$ declines, and $L_{\mathrm{grav}}^{(\mathrm{nh})}$
increases. The positive contribution of $L_{\mathrm{grav}}^{(\mathrm{h})}$
due to the response of the radiative layers dominates [panel (d),
Fig.~\ref{Legravs2}], causing a rise in $L_{2}$ and an expansion of the
surface layers [e.g. point (d), Figs.~\ref{rad_lum_01} and \ref{rad1}].

\begin{figure}
  \begin{center}
    \includegraphics[trim = 0mm 0mm 0mm
    5mm,clip,scale=0.45]{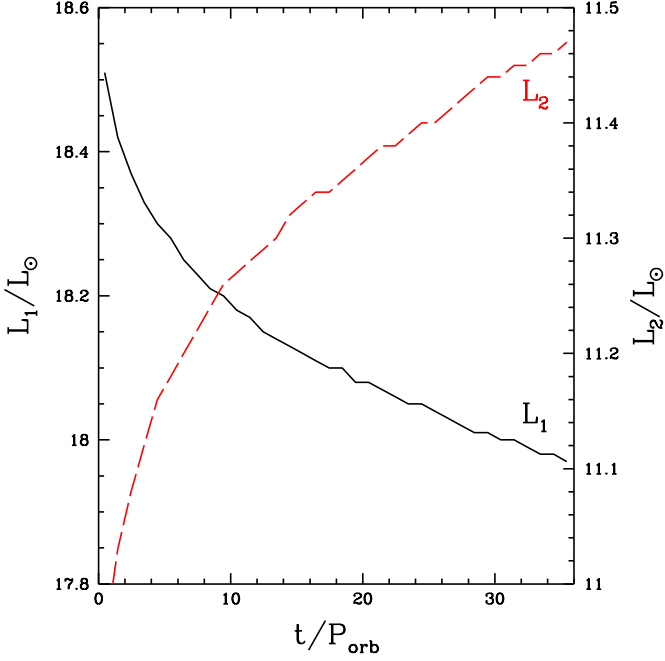}
    \caption{Evolution of the surface luminosity of the donor $L_{1}$
      (solid black curve, left axis) and of the accretor $L_{2}$ (dashed
      red curve, right axis) just before mass transfer recommences over a
      duration of about 30 periastron passages.}
  \label{L1_L2} 
  \end{center}
\end{figure}

\begin{figure}
  \begin{center}
    \includegraphics[trim = 0mm 0mm 0mm
    5mm,clip,scale=0.46]{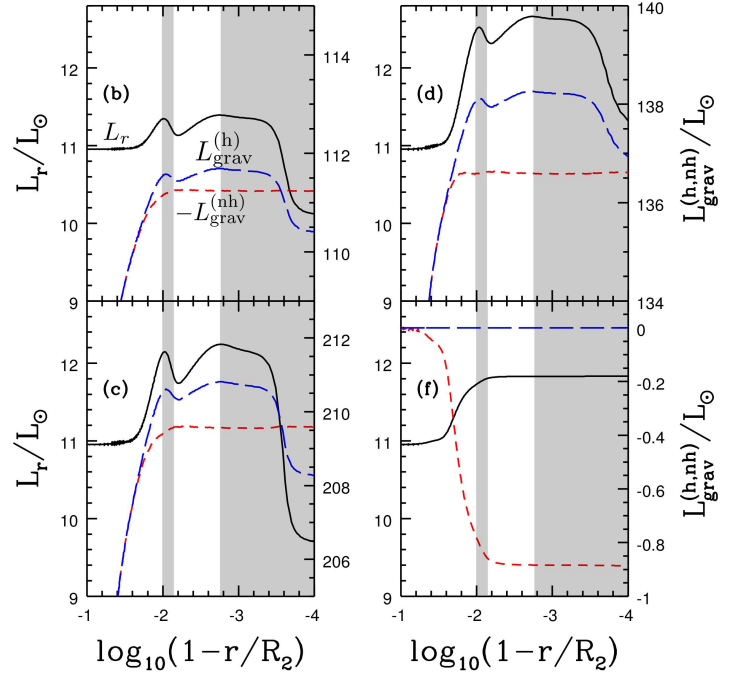}
    \caption{Similar to Fig.~\ref{Legravs} but now for the accretor. Note
      that $-L_{\mathrm{grav}}^{(\mathrm{nh})}$ has been plotted (right
      axis).}
    \label{Legravs2}
  \end{center}
\end{figure}

At $t \geq 0.53P_{\mathrm{orb}}$ since apastron, mass accretion has become
negligible ($L_{\mathrm{grav}}^{(\mathrm{h})}\approx 0$) and contraction
resumes [$L_{\mathrm{grav}}^{(\mathrm{nh})}>0$; see panel (f),
Fig.~\ref{Legravs2}] as the excess energy originally created in the
radiative layers by the deposition of matter is now radiated away. Note
that $|L_{\mathrm{grav}}^{(\mathrm{nh})}|$ is much smaller [panel (f)]
because the accretor is no longer perturbed by mass deposition, and is
relaxing towards thermal equilibrium. Hence, $L_{2}$ and $R_{2}$ decrease
slightly [point (f) in Figs. \ref{rad_lum_01} and \ref{rad1}]. Subsequent
mass transfer episodes causes $L_{2}$ and $R_{2}$ to gradually rise from
orbit to orbit (red curve in Fig.~\ref{L1_L2}), again in accordance with
studies of accreting radiative stars in circular binaries (e.g. Neo et
al. \cite{neo1977}, Webbink \cite{webbink76}).

The impact of tides and rotation on the accretor is the same as for the
donor, except that at $t\ga0.52P_{\mathrm{orb}}$, the distorted model has a
somewhat higher luminosity than the non-distorted case. This is because the
higher mass accretion rate creates a larger energy excess in the radiative
layers below the surface convection zone due to compression.

For the non-distorted models, we find that, during the periastron passage,
the surface luminosity of the donor decreases by as much as
$\Delta{L}_{1}\approx 1.3$ $L_{\odot}$, while for the accretor it increases
by as much as $\Delta{L}_{2}\approx 2$ $L_{\odot}$. However, these
variations almost cancel out, and the net change in the luminosity for the
entire system ($\Delta{L}_{\Sigma}=\Delta{L}_{1}+\Delta{L}_{2}$) is
$\Delta{L}_{\Sigma}\approx 0.7$ $L_{\odot}$. This corresponds to a
variation of about 0.03 magnitude, which may not be easily observable. For
the distorted models, on the other hand, we find $\Delta{L}_{1}\approx
-2.6$ $L_{\odot}$ and $\Delta{L}_{2}\approx 6.5$, and so a change of about
0.1 magnitude ($\Delta{L}_{\Sigma}\approx 3.9$ $L_{\odot}$), which
detection may be more feasible. We emphasize however that despite the small
values of $\Delta L_{\Sigma}$, the change in the intrinsic luminosities of
each star is significant and takes places over a brief period of time
($\approx$ 1-2 hours in our model).

\subsection{Effect of direct impact accretion}
\label{sec:accretion}

For an accretion disc to form, the minimum distance of approach between the
matter stream and the accretor must satisfy
\begin{equation}
  R_{\mathrm{min}}\approx 0.0425D[q(1+q)]^{1/4}>R_{2},
\label{R_min}
\end{equation}
(Ulrich \& Burger \cite{ulrich}). For our system, $R_{\mathrm{min}}\approx
0.2$ R$_{\odot}<R_{2}$, implying direct impact accretion and the formation
of a hot-spot, in accord with LS11. The presence of a hot-spot may give
rise to mass ejection from the system. To investigate this possibility, we
follow van Rensbergen et al. (\cite{rensbergen_2008}), and briefly
summarise below their formalism that we implemented within BINSTAR.

According to these authors, the hot-spot luminosity, $L_{\mathrm{HS}}$, is
given by
\begin{equation}
  L_{\mathrm{HS}}= \frac{\alpha_{\mathrm{acc}}}{S_{\mathrm{acc}}}L_{\mathrm{acc}}=KL_{\mathrm{acc}},
\label{L_hs}
\end{equation}
where $S_{\mathrm{acc}}$ is the fraction of the accretor's surface
inhabited by the hot-spot and $\alpha_{\mathrm{acc}}$ is the accretion
efficiency. A fit to the variable $K$ was performed by van Rensbergen et
al. (\cite{rensbergen_2011}) using an observed sample of 13 Algol binaries,
and is given by
\begin{equation}
  K\approx 3.9188\left(\frac{M_{1}+M_{2}}{M_{\odot}}\right)^{1.645}.
  \label{K}
\end{equation}
The accretion luminosity, $L_{\mathrm{acc}}$, is calculated from the
potential difference between the \l1 point, $\Psi(x_{\L1},0,0)$ and the
point of impact, $\Psi_{\mathrm{imp}}(x_{\mathrm{imp}},y_{\mathrm{imp}},0)$
(using Eq. \ref{Psi}), yielding
\begin{equation}
  L_{\mathrm{acc}}=|\dot{M}_{1}|(\Psi_{\L1}-\Psi_{\mathrm{imp}}).
\label{L_acc}
\end{equation}
The point of impact, $(x_{\mathrm{imp}},y_{\mathrm{imp}})$, can be found
from ballistic trajectory calculations. However, to do this for eccentric
systems where the separation, and hence the potential, are varying with
time, is beyond the scope of the paper. To estimate
$(x_{\mathrm{imp}},y_{\mathrm{imp}})$, we use the angle between the point
of impact and the line joining the two stars, $\varpi$, determined from the
ballistic trajectories calculated by Flannery (\cite{flannery}) for
different values of $q$ in circular orbits. For our system, we use
$\varpi\approx 70^{\circ}$. From Eqs. (\ref{L_hs}) and (\ref{L_acc}), it is
obvious that $L_{\mathrm{HS}}$ follows the same orbital modulation as
$\dot{M}_{1}$, and if $\dot{M}_{1}$ exceeds some critical value
$\dot{M}_{\mathrm{crit}}$, the luminosity in the hot-spot region may exceed
the Eddington value, $L_{\mathrm{Edd}}$, allowing for mass ejection. van
Rensbergen et al. (\cite{rensbergen_2008}) give
\begin{dmath}
  \dot{M}_{\mathrm{crit}} = \frac{1}{\Psi_{\L1}-\Psi_{\mathrm{imp}}}
  \left\{\left[\frac{L_{\mathrm{edd}}R_{2}}{GM_{2}}\left(\frac{GM_{1}}{r_{1,\mathrm{imp}}}+\frac{GM_{2}}{R_{2}}+\frac{1}{2}r_{\mathrm{C,imp}}^{2}\omega^{2}\\-\frac{1}{2}R_{2}^{2}\Omega^{2}_{2}\right)
      - L_{2}\right]\frac{1}{K}+\dot{E}_{\mathrm{rot}}\right\},
\label{M_crit}
\end{dmath}
where $r_{1,\mathrm{imp}}$ and $r_{\mathrm{C,imp}}$ are the distances from
the point of impact to the centre of mass of the donor and the binary
system respectively and $\dot{E}_{\mathrm{rot}}$ is the rate of change of
the accretor's rotational kinetic energy. Due to the small amount of mass
that is transferred (up to approximately $10^{-7}$ $M_{\odot}$ in one
orbit), the change in the accretor's spin speed is negligible over the
timescale of the simulation, and $\dot{E}_{\mathrm{rot}}\approx 0$.

The temperature of the hot-spot, $T_{\mathrm{HS}}$, can be estimated from
$\sigma T^{4}_{\mathrm{HS}}=L_{\mathrm{HS}}/S_{\mathrm{HS}}$, where
$S_{\mathrm{HS}}\approx 10^{20}$ cm$^{2}$ is the area of the hot-spot on
our accretor's surface (see Gunn et al. \cite{gunn}, van Rensbergen et
al. \cite{rensbergen_2008}). At periastron, $T_{\mathrm{HS}}\approx 10^{5}$
K, so the material in that region is fully ionized, justifying our use of
the electron scattering opacity in the expression for $L_{\mathrm{edd}}$.

\subsection{Effects of asynchronism and eccentricity}
\label{sub:asynch}

\begin{figure}
  \begin{center}
    \includegraphics[trim = 0mm 0mm 0mm
      5mm,clip,scale=0.40]{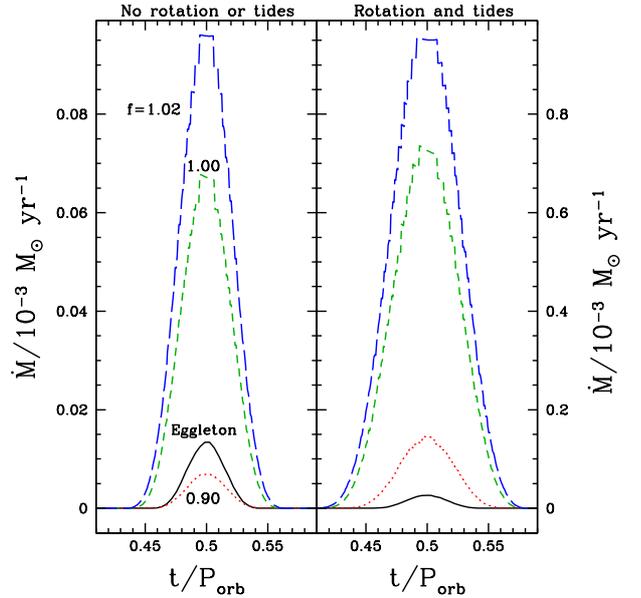}
      \caption{Left panel, left axis: similar to Fig.~\ref{mtransfer_01},
        but here the effects of asynchronism and eccentricity on the Roche
        lobe radius are taken into account (Eqs. \ref{Psi} and
        \ref{async}). Black, solid curve: standard Roche lobe formalism;
        red, dotted curve: $f=0.90$; green, short-dashed curve: $f=1.00$;
        blue, long-dashed curve: $f=1.02$. Right panel, right axis: the
        same but for the distorted models.}
      \label{mdots_async}
    \end{center}
\end{figure}

\begin{figure}
  \begin{center}
    \includegraphics[trim = 0mm 0mm 0mm
    10mm,clip,scale=0.40]{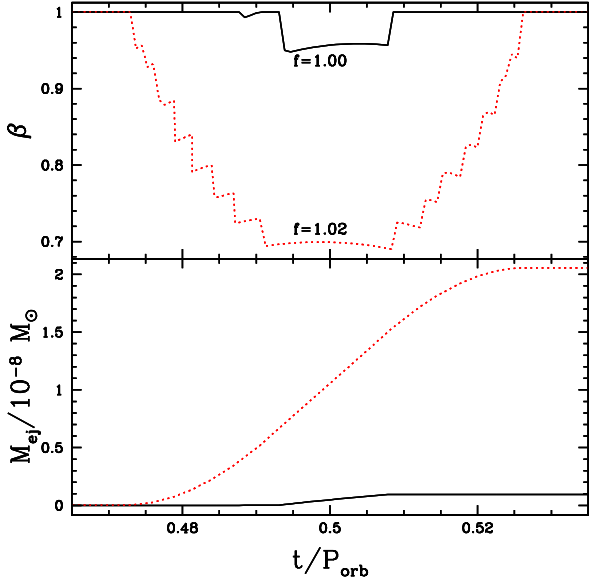}
    \caption{Upper panel: Accretion efficiency,
      $\beta=|\dot{M}_{2}/\dot{M}_{1}|$ versus time for the distorted
      models, where $R_{\L1}$ is determined from Eqs. (\ref{Psi}) and
      (\ref{async}), with $f=1.00$ (black, solid curve) and $f=1.02$ (red,
      dotted curve). Lower panel: mass of ejected material,
      $M_{\mathrm{ej}}$ (in units of $10^{-8}$ $M_{\odot}$), as a function
      of time.}
    \label{beta_RLOF}
  \end{center}
\end{figure}

Fig.~\ref{mdots_async} shows the effects on $\dot{M}_{1}$ of an eccentric
orbit and an asynchronously rotating donor on the Roche lobe
radius\footnote{The step-like features near periastron, particularly
  evident for the synchronous and super-synchronous models, are a result of
  low spatial resolution in the surface layers of the donor star, revealed
  by the small time-steps used. This produces the corresponding features in
  Fig.~\ref{beta_RLOF}}(Eqs. \ref{Psi} and \ref{async}). Three cases are
considered: sub-synchronous rotation of the donor with the orbital motion
at periastron, $f=0.90$ (green, short-dashed curves), synchronous rotation
($f=1.00$, red, dotted curves) and super-synchronous rotation ($f=1.02$,
blue, long-dashed curves).

\begin{figure}
  \begin{center}
    \includegraphics[trim = 0mm 0mm 0mm
    5mm,clip,scale=0.40]{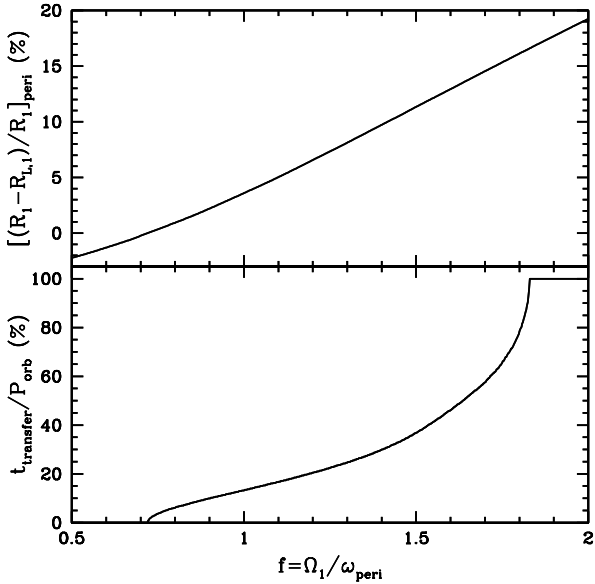}
    \caption{Relative difference between the donor radius and its Roche
      lobe radius at periastron, $[(R_{1}-R_{\L1})/R_{1}]_{\mathrm{peri}}$
      (top panel), and the duration of mass transfer,
      $t_{\mathrm{transfer}}$, in units of the orbital period (bottom
      panel), as a function of $f$ for our 1.50 $M_{\odot}$ donor.}
      \label{RL_async}
    \end{center}
\end{figure}

For the non-distorted models, at a given orbital phase, increasing the
value of $f$ increases $\dot{M}_{1}$. At periastron, for example,
$\dot{M}_{1}$ rises from approximately $7.0\times{10}^{-6}$ $M_{\odot}$
yr$^{-1}$ for the sub-synchronous case to about $9.6\times{10}^{-5}$
$M_{\odot}$ yr$^{-1}$ for the super-synchronous case; about a factor of 14
increase. The explanation resides in the fact that a higher value of $f$
yields a smaller Roche lobe radius. In turn, the donor will over-fill its
Roche lobe further (i.e. an increase in $(R_{1}-R_{\L1})/R_{1}$ with
increasing $f$; see left panel of Fig.~\ref{RL_async}, red curve), which
will cause the mass transfer rate to grow (Eqs. \ref{mdot_ritter} and
\ref{mdot_kolb}). Despite the enhanced mass loss rate, we find that mass
transfer is still fully conservative. Values of $\dot{M}_{\mathrm{crit}}$
are higher than $\dot{M}_{1}$ at periastron by about a factor of 100 for
the sub-synchronous case to about a factor of 6 for the super-synchronous
case.

On the other hand, for the tidally and rotationally distorted models, the
mass transfer rate is typically a factor of 10 higher than for the
non-distorted configuration, for a given value of $f$. This increase in
$\dot{M}_{1}$ leads to both a higher hot-spot luminosity (as high as about
$10^{5}$ $L_{\odot}$ at periastron for the distorted, super-synchronous
case) and non-conservative evolution for the synchronous and
super-synchronous models (left panel, Fig.~\ref{beta_RLOF}), with values of
$\dot{M}_{\mathrm{crit}}$ of $7\times{10}^{-4}$ $M_{\odot}$ yr$^{-1}$ and
$6\times{10}^{-4}$ $M_{\odot}$ yr$^{-1}$, respectively. The accretion
efficiency $\beta$ at a given phase is smaller for the super-synchronous
case, because for this model $\dot{M}_{1}$ is higher, and so it further
exceeds $\dot{M}_{\mathrm{crit}}$ leading to more mass ejection. Due to the
modulation of $\dot{M}_{1}$ with orbital phase, $\beta$ is a minimum at
periastron. During one orbit, the total ejected mass from the accretor is
about ${10}^{-9}$ $M_{\odot}$ and $2\times{10}^{-8}$ $M_{\odot}$ for the
synchronous and sub-synchronous models, respectively.

The change in the Roche lobe radius with $f$ can be explained in terms of
the location of the $\mathcal{L}_{1}$ point. Recall that $\mathcal{L}_{1}$
is where $\partial \Psi/\partial x=0$, i.e. where a particle experiences no
net acceleration, due to the balance of the centrifugal and gravitational
accelerations. Let the position of the $\mathcal{L}_{1}$ point at a given
phase for a given value of $f$ be $x_{\mathcal{L}_{1}}(0)$. If $f$ is then
increased (causing a corresponding increase in $\mathcal{A}$), then a test
particle at $x_{\mathcal{L}_{1}}(0)$ will experience a stronger centrifugal
acceleration, and hence an outward displacement. The new location of
$x_{\L1}$ therefore needs to be situated closer to the donor star
[$x_{\L1}<x_{\L1}(0)$] to re-establish a net zero-acceleration. Since the
Roche equipotential surface passes through the $\mathcal{L}_{1}$ point, a
decreasing value of $x_{\L1}$ means that both the volume and the radius of
the Roche lobe will shrink.

A shrinking Roche lobe radius with rising $f$ also means that mass transfer
will occur for a longer duration, because mass transfer will start (end)
earlier (later) in the orbit. This duration, $t_{\mathrm{transfer}}$, is
shown in the right panel of Fig.~\ref{RL_async}. For our super-synchronous
case, mass transfer lasts for approximately 17 per cent of the orbital
period.

We also see that if $f\la 0.7$, mass transfer does not occur at all. This
is because the Roche lobe radius of the sub-synchronously rotating donor
increases so much, compared to the value obtained from Eq. (\ref{R_L_egg}),
that the donor star never fills it. Indeed, as Fig. \ref{RL_async} shows,
$(R_{1}-R_{\L1})/R_{1}<0$ at periastron for $f\la 0.7$.  Therefore, if we
evolve our 1.50+1.40 $M_{\odot}$ system with $f=0.36$, our system will
never undergo mass transfer, in contrast with LS11. At periastron, for
$f=0.36$, the relative increase in the Roche lobe radius due to
asynchronous rotation, compared to the value obtained from
Eq. (\ref{R_L_egg}), is approximately 6 per cent, in agreement with
Sepinsky et al. (\cite{sepinsky_07a}). We return to the comparison of our
work with that of LS11 in Section \ref{section:discussion}. On the other
hand, for $f\ga 1.8$, Fig.~\ref{RL_async} shows that mass transfer will
occur over the whole orbit, since at all phases
$R_{\mathcal{L}_{1}}<R_{1}$.

Finally, the change in $\mathcal{A}$ with orbital phase, will cause
a corresponding change in the donor's Roche lobe radius. To understand this
more clearly, we can recast Eq. (\ref{async}) in terms of the instantaneous
ratio of the spin angular speed of the donor star, and the orbital angular
velocity, i.e.
\begin{equation}
  \mathcal{A}=\left(\frac{\Omega_{1}}{\omega}\right)^{2}(1+e\,\mathrm{cos}\nu).
\label{async2}
\end{equation}
As we move from periastron to apastron, the orbital speed
decreases. Therefore, for a given value of $\Omega_{1}$, the donor star
will rotate progressively more super-synchronously with the orbital motion,
causing a corresponding rise in both $\mathcal{A}$, and the centrifugal
acceleration. Consequently, the Roche lobe radius will be progressively
smaller than the value calculated by the standard Eggleton
(\cite{eggleton}) formalism, as apastron is approached.

\section{Discussion}
\label{section:discussion}


The evolution of the mass transfer rate has a Gaussian-like behaviour, with
a maximum value occurring at periastron, in excellent agreement with the
findings of LS11. However, we also report major differences.

LS11 obtain a maximum mass transfer rate which occurs in fact slightly
after periastron. This delay is due to the time required by the material
ejected from the donor to fall down the potential well of the companion
star and be accreted. This is approximately the free-fall time, of the
order of $\tau_{\mathrm{ff}}\approx 0.06P_{\mathrm{orb}}$ (LS11). In
BINSTAR, this delay is absent because the code does not currently follow
the ballistic trajectory of the ejected material. Instead, it assumes that
mass transfer occurs instantaneously. The delay found by LS11 is more
physical as a result of their more realistic treatment of mass transfer,
but it does not change the overall picture.

Furthermore, we emphasize that our determination of the point of impact on
the accretor's surface based on the angle $\varpi$ is rather
crude. Sepinsky et al. (\cite{sepinsky_10}) calculated ballistic
trajectories in eccentric binaries for a single particle ejected at
periastron. They found that the particle may fall back onto the donor star,
as well as fall onto the accretor via a disc or direct impact. However, as
discussed below, mass transfer occurs over a significant fraction of the
orbital period, and not just at periastron. Furthermore, their study
neglects the contribution of the thermal sound speed at the \l1 point to
the ejected particle's velocity. It is unclear how these modifications
would impact on the ballistic trajectory and on the mass transfer.

If we take into account the effects of asynchronous rotation of the donor
(with $f\approx 0.36$) and the eccentricity of the orbit on the Roche lobe
radius, our donor never fills its Roche lobe, even if the effect of tides,
which act to increase the donor radius, are considered. This is in contrast
to LS11, who find that mass transfer lasts for approximately 25 per cent of
the orbital period, and peaks at periastron at a value of
$\dot{M}_{1}\approx 2\times10^{-3}$ $M_{\odot}$ yr$^{-1}$.  In their
simulation, tidal and rotational forces increase the radius of their donor
to a greater extent, to approximately 1.8 $R_{\odot}$, so that the donor is
sufficiently large to fill its Roche lobe near periastron. As a check, we
re-ran a simulation using a slightly more evolved donor star with
$R_{1}\approx 1.8$ $R_{\odot}$, again accounting for the effects of
asynchronism and eccentricity on the value of $R_{\L1}$. Now, we find that
mass transfer lasts for about 30 per cent of the orbital period, in good
agreement with LS11, but our peak mass transfer rate is approximately a
factor of 35 larger than theirs. This may arise from the uncertainty in
$R_{1}$, since LS11 cannot provide a precise value. Even though this
uncertainty is small, it may greatly impact upon $\dot{M}_{1}$ due to the
sensitivity on the degree of overflow (Eqs. \ref{mdot_ritter} and
\ref{mdot_kolb}).

The technique employed by LS11 to determine mass transfer rates is
obviously more physically realistic than the 1-dimensional analytical
scheme used in the present study. The assumptions underlying
Eqs. (\ref{mdot_ritter}) and (\ref{mdot_kolb}) that the flow is laminar,
and that material moves along the equipotential surfaces is not strictly
valid. Instead, material flows upwards from within the Roche lobe of the
donor, rather than along its surface (Eggleton \cite{eggleton_2006}), which
would have the effect of increasing the mass transfer rate. On the other
hand, the pressure gradients will produce turbulence in the flow rather
than bulk motion of the material, and will contribute to reduce the mass
transfer rate. Ge et al. (\cite{ge10}), who use a similar formalism to
Eq. (\ref{mdot_kolb}), estimate that the calculated mass transfer rate is
accurate to within about a factor of 2.

Despite the fact that LS11 distribute their SPH particles according to
density profiles calculated from their stellar evolution code, they do not
consider the transport of energy due to convection and radiation,
(similarly to SPH simulations by Reg\H{o}s et al. \cite{regos} and Church
et al. \cite{church}). Instead, LS11 apply a simple polytropic equation of
state of the form
\begin{equation}
P=\rho(\gamma-1)u,
\label{P_lajoie}
\end{equation}
where $u$ is the specific internal energy of the stellar material, and
$\gamma=5/3$. While a value of $\gamma=5/3$ is a reasonable approximation
for the deep optically thick layers of the interior, it is inadequate to
treat the regions of partial ionization or small optical
thickness. Considering that the layers of the star in LS11 are adiabatic,
their SPH simulations will over-estimate the pressure and the temperature
in these layers, yielding larger radii. This `relaxed' SPH stellar
configuration will be able to expand to a greater extent due to tidal and
rotational deformation, and impact on the mass transfer rate.

Even though our simulations are able to provide a more realistic internal
structure, the treatment of tides and rotation outlined in Section
\ref{sec:tides} remains an approximation, since all 3-dimensional effects
are folded into a 1-dimensional formalism (see also Knigge, Baraffe \&
Patterson \cite{knigge} for a discussion of this matter). It is therefore
unclear whether the discrepancy in stellar radii, and therefore the mass
transfer rate, is due to the polytropic approximation used by LS11 or the
subtleties of the Kippenhahn \& Thomas (\cite{kippenhahn_76}) method used
here.

For our distorted, super-synchronous model, which gives a similar mass
transfer rate at periastron as found by LS11, mass transfer is
non-conservative. Over one orbit, the total mass lost by the donor is
approximately $1\times{10}^{-7}$ $M_{\odot}$, while the total accreted mass
is about $8\times{10}^{-8}$ $M_{\odot}$, i.e. 20 per cent of the
transferred material is ejected by the accretor. However, in contrast to
SPH simulations, we cannot follow the ultimate fate of this material. For
their 1.50+1.40 $M_{\odot}$ system, 20 per cent of the transferred material
formed an envelope around the system, while 5 per cent was ejected. On the
other hand, since LS11 do not model the thermal structure of the hot-spot,
they may under-estimate the fraction of material that is ejected from the
system.

It is evident from Figs. \ref{mdots_async} and \ref{RL_async} that the
degree of asynchronism of the donor star has a significant impact on both
the mass loss rate and the duration of mass transfer. It should be pointed
out that the mass transfer rates predicted by the standard Eggleton
(\cite{eggleton}) formalism (which assumes that the star is rotating
synchronously with the orbit at each location) and the $f=1.00$ case with
non-zero eccentricity differ by approximately a factor of about 5. This
factor is further enhanced to about 60 due to tidal and rotational effects
on the donor star. Hence, it is imperative that the effect of eccentricity
and asynchronism on the Roche lobe geometry, and the distortion of the
donor star, be accounted for.

In the Sepinsky et al. (\cite{sepinsky_07b, sepinsky_09}) studies, the
structure and the response of the donor star were not taken into account,
and mass transfer was modelled as a delta function at periastron. This is
in contrast with Lajoie \& Sills (\cite{lajoie_11a}), LS11 and the present
work which show that the episode of mass exchange represents a
non-negligible fraction of the orbital period. Allowing for
non-instantaneous mass transfer may impact on the evolution of the binary
system in a way that is different to the delta function model.

\section{Summary}
\label{section:conclusions}

This study paves the first steps towards calculating mass transfer rates
for significantly eccentric binaries, taking into account the effects of
tides and rotation on the structures of the stellar components, as well as
the effects of eccentricity and asychronous rotation of the donor on the
Roche lobe radius. In this paper, we have used a state-of-the-art binary
evolution code BINSTAR to calculate mass transfer rates for a 1.50+1.40
$M_{\odot}$ main sequence binary system, with an eccentricity of 0.25 and
an orbital period of $P_{\mathrm{orb}}\approx 0.7$d and we compared our
results with the SPH calculations performed by LS11 for the same system.

The evolution of the mass transfer rate with time shows Gaussian-like
behaviour, with a maximum mass transfer rate occurring at periastron, in
qualitative agreement with Lajoie \& Sills (\cite{lajoie_11b}). The
accretion luminosity (which emanates from a hotspot for this particular
system) also varies in response to the changing mass transfer rate. The
duration of the mass transfer rate represents a non-negligible fraction of
the orbital period, particularly if the donor star is rotating sufficiently
rapidly. On the other hand, mass transfer may not occur at all if the donor
is rotating too slowly.

During RLOF, the timescale over which the mass transfer rate changes,
$\tau_{\dot{M}}=|\dot{M}_{1}/\ddot{M}_{1}|$ is so short that only the
outer-most stellar layers, consisting of the surface convection zone, can
restore thermal equilibrium. Initially, the response of each star is
dictated by this convection zone. As mass transfer proceeds, a larger
fraction of the stars' envelope, consisting of the radiative layers,
dictate the subsequent response of each star.

The evolution of the luminosity of the donor and accretor over 30 orbits
(Fig \ref{L1_L2}) highlights the fact that the longer-term evolution of the
stellar structure is what we would expect for stars with significantly
radiative envelopes, as shown by studies of such stars in circular orbits.

Finally, tidal and rotational forces increase the surface radius of the
donor, enhancing mass transfer further by a factor of about 10, potentially
leading to non-conservative mass transfer.

In the future, we wish to follow the mass transfer rate as a function of
orbital phase over many orbits, and determine the secular orbital evolution
of eccentric binaries.

\begin{acknowledgements}
  PJD acknowledges financial support from the Communaut\'{e} Fran\c{c}aise
  de Belgique - Actions de recherche Concert\'{e}es, from the
  Universit\'{e} Libre de Bruxelles and from an FNRS Fellowship. PJD would
  also like to thank U. Kolb and J. Sepinsky for many useful discussions,
  and A. Sills for providing further details of the stellar models used in
  the SPH simulations. We also thank S. Toupin for his assistance with the
  development of the BINSTAR routines which calculates the Roche lobe
  radius, and the distortion of the star due to tides and
  rotation. Finally, we are grateful to the anonymous referee whose
  constructive comments improved the quality of the manuscript. LS is a
  FNRS research associate.
\end{acknowledgements}

\begin{appendix}

\section{Determining the Function $F^{*}(q,\mathcal{A})$}
\label{section:m_transfer} 

In the following description, we use a coordinate system where the origin
lies at the centre of mass of the donor. The positive $x$-axis points from
the centre of mass of the donor to the centre of mass of the accretor,
while the $z$-axis is perpendicular to the orbital plane, and is parallel
to the donor's spin angular velocity. The $y$-axis lies on the orbital
plane, and forms a right-handed set. 

We start with Eq. (20) of Sepinsky et al. (\cite{sepinsky_07a}), which
gives the potential, $\Psi$ (normalised by the potential due to the
accretor, $\frac{GM_2}{D}$)
\begin{eqnarray}
\Psi=-\frac{q}{(x^{2}+y^{2}+z^{2})^{\frac{1}{2}}}-\frac{1}{[(x-1)^{2}+y^{2}+z^{2}]^{\frac{1}{2}}}
\nonumber \\
-\frac{1}{2}\mathcal{A}(1+q)(x^{2}+y^{2})+x,
\label{Psi_app}
\end{eqnarray}
where all coordinates are given in units of the instantaneous separation,
$D$ and $\mathcal{A}$ is given by Eq. (\ref{async}). Following Meyer \&
Meyer-Hofmeister (1983) and Ritter (1988), we can express the change in the
potential around the vicinity of the $\mathcal{L}_1$ point, $\Delta \Psi$,
using a Taylor expansion to the second order, according to
\begin{equation}
\Delta\Psi\approx By^2+Cz^2,
\label{DeltaPsi}
\end{equation}
where
\begin{equation}
B=\frac{1}{2}\left(\frac{\partial^{2}\Psi}{\partial y^{2}}\right)_{\mathcal{L}_{1}}=\frac{1}{2}[g(q)-\mathcal{A}-q\mathcal{A}],
\label{ddpsi_dyy} 
\end{equation}
where the subscript $\mathcal{L}_{1}$ indicates that the derivative is
evaluated at the inner-Lagrangian point, and $g(q)$ is given by
Eq. (\ref{gq}). Similarly,
\begin{equation}
C=\frac{1}{2}\left(\frac{\partial^{2}\Psi}{\partial z^{2}}\right)_{\mathcal{L}_{1}}=\frac{1}{2}g(q).
\label{ddpsi_dzz}
\end{equation}
Recalling that the potential has been expressed in units of the
secondary's gravitational potential, and that the distances are in
units of the separation, then converting back to `normal' units gives
us for $B$ and $C$, respectively,
\begin{eqnarray}
B & = & \frac{GM_{2}}{2D^3}[g(q)-\mathcal{A}-q\mathcal{A}] \nonumber \\
  & = & \frac{GM_{1}}{2qD^3}[g(q)-\mathcal{A}-q\mathcal{A}],
\label{B_2}
\end{eqnarray}

\begin{equation}
C=\frac{GM_{1}}{2qD^3}g(q)
\label{C_2}
\end{equation}
We can see from Eq.(\ref{DeltaPsi}) that the equipotential surface in the
plane of the $\mathcal{L}_1$ point is an ellipse with an area, $S$, given
by
\begin{equation}
S=\pi \frac{\Delta \Psi}{\sqrt{BC}}.
\label{area}
\end{equation}
Following Meyer \& Meyer-Hofmeister (1983), the area of the cross section of the
flow, $Q$, is then
\begin{equation}
Q=\pi\frac{\mathcal{R}T}{\mu}\frac{1}{\sqrt{BC}}=\frac{2\pi\mathcal{R}T}{\mu}\frac{D^{3}q}{GM_{1}}\left\{
  g(q)[g(q)-\mathcal{A}-q\mathcal{A}]\right\} ^{-\frac{1}{2}}.
\label{Q}
\end{equation}
Defining
\begin{equation}
  \beta=\frac{R_{\L1}}{D},
\label{beta_app}
\end{equation}
then substituting Eq.(\ref{beta_app}) into (\ref{Q})
gives us
\begin{equation}
Q=\frac{2\pi\mathcal{R}T}{\mu}\frac{R_{\L1}^{3}}{GM_{1}}
  q\,\beta^{-3}\,\left\{g(q)[g(q)-\mathcal{A}-q\mathcal{A}]\right\} ^{-\frac{1}{2}}
\label{Q_2}
\end{equation}
The mass transfer rate when the donor star exactly fills its Roche
lobe, $\dot{M}_{0}$, is given by (Ritter \cite{ritter})
\begin{equation}
\dot{M}_{0}=\frac{1}{\sqrt{e}}\ \rho_{\mathrm{ph},1}\,v_{\mathrm{s}}\,Q,
\label{Mdot_app}
\end{equation} 
where
\begin{equation}
v_{\mathrm{s}}=\frac{\mathcal{R}T_{\mathrm{eff},1}}{\mu_{\mathrm{eff},1}},
\label{vs}
\end{equation}
is the sound velocity of the material in the location of the
$\mathcal{L}_1$ point. Substituting Eqs. (\ref{Q_2}) and (\ref{vs}) into
(\ref{Mdot_app}) and comparing the result with Eq. (\ref{mdot_0}) gives us
\begin{eqnarray}
\dot{M}_{0} & = &
\frac{2\pi}{\sqrt{e}}q\beta^{-3}\left\{g(q)[g(q)-\mathcal{A}q-\mathcal{A}]\right\}^{-1/2}
\nonumber \\
            &   &
            \times\frac{R_{\mathrm{L}}^3}{GM_{1}}\left(\frac{\mathcal{R}T_{\mathrm{eff}}}{\mu_{\mathrm{ph,1}}}\right)^{3/2}\rho_{\mathrm{ph,1}},
            \nonumber \\
            & = & \frac{2\pi}{\sqrt{e}}F^{\star}(\mathcal{A},q)\frac{R_{\mathrm{L}}^3}{GM_{1}}\left(\frac{\mathcal{R}T_{\mathrm{eff}}}{\mu_{\mathrm{ph,1}}}\right)^{3/2}\rho_{\mathrm{ph,1}},
\label{mdot_0_appendix}
\end{eqnarray}
where 
\begin{equation}
F^{*}(A,q)=q\beta^{-3}\left\{g(q)[g(q)-\mathcal{A}q-\mathcal{A}]\right\}^{-1/2},
\label{F_star_appendix}
\end{equation}
as given by Eq. (\ref{F_star}).

Finally, the pressure scale scale height at the location of the
$\mathcal{L}_{1}$ point is calculated using
\begin{equation}
\hat{H}_{\mathrm{P}}=\frac{H_{\mathrm{P}}}{\gamma(q)},
\label{HP_RL}
\end{equation}
where $\gamma(q)$ takes into account the non-spherical shape of the donor's
Roche lobe, and is given by (Ritter \cite{ritter})
\[\gamma(q)=\left\{
\begin{array}{l l}
0.954+0.025\,\mathrm{log}\,q-0.038(\mathrm{log_{10}}\,q)^{2} & \quad \mbox{if
  $0.04\leq q < 1$}\\
0.954+0.039\,\mathrm{log}\,q-0.114(\mathrm{log_{10}}\,q)^{2} & \quad \mbox{if
  $1\leq q < 20$} \\
\end{array} \right. \] 


\end{appendix}

\end{document}